\documentclass[preprint,12pt]{elsarticle}

 \usepackage{epsfig}

\usepackage{amssymb}
\newcommand{\ImagehereCaption}[3]{\begin{figure}[h!]

\begin{center}
\centerline{\psfig{file=#1,width=#2}}
 
 \caption{#3}
\end{center}
\end{figure}
}
 \newcommand{\ImagehereFull}[4]{\begin{figure}[hbt]
\centerline{\psfig{file=#1,width=#2}}
\caption{#3}
    \label{#4}
\end{figure}
}
\newcommand{\barra}[1]{\overline{\hspace{-1pt} #1 \hspace{-1pt} }}

\usepackage{ulem}
\newcommand{\Not}[1]{\mbox{\scriptsize{\sout{\ensuremath{#1}}}}}

\journal{Theoretical Biology}

\begin{document}

\begin{frontmatter}

\title{Is modularity the reason why recombination is so ubiquitous?}

\author[C3,ICN]{Manuel\ Beltr\'an del R\'\i o \corref{cor1}}
\ead{manuel.beltrandelrio@nucleares.unam.mx}
\cortext[cor1]{Corresponding Author}

\author[C3,ICN]{Christopher R.\ Stephens}
\author[C3,IIMATE]{David A. \ Rosenblueth}

\address[C3]{$C_3$ - Centro de Ciencias de la Complejidad, M\'exico D.F.}

\address[ICN]{Instituto de Ciencias Nucleares, UNAM\\
Circuito Exterior, A.\ Postal 70-543, M\'exico D.F.\ 04510}

\address[IIMATE]{Instituto de Investigaciones en Matem\'aticas Aplicadas y en Sistemas, UNAM\\
A.\ Postal 20-726, M\'exico D.F.\ 01000}
\begin{abstract}
Homologous recombination is an important operator in the evolution of biological organisms. 
However, there is still no clear, generally accepted understanding
of why it exists and under what circumstances it is useful. In this paper we
consider its utility in the context of an infinite population
haploid model with selection and homologous recombination. We define utility in terms
of two metrics - the increase in frequency of fit genotypes, and the increase in average 
population fitness, relative to those associated with selection only. 
Explicitly, we explore the full parameter space of a two-locus
two-allele system, showing, as a function of the landscape and the initial population, that
recombination is beneficial in terms of these metrics in two distinct regimes: a relatively landscape
independent regime -  the {\it search} regime - where recombination aids in the search for a fit 
genotype that is absent or at low frequency in the population; and the {\it modular} regime, 
where recombination allows for the juxtaposition of fit ``modules" or Building Blocks. Thus, 
we conclude that the ubiquity and utility of recombination is intimately associated with the 
existence of modularity and redundancy in biological fitness landscapes.

\end{abstract}


\end{frontmatter}

\section{Introduction}
\label{Intro}

The existence, prevalence and utility of genetic recombination is an old and enduring puzzle of biology \cite{maynard1971use}. 
Seminal works, such as \cite{FeldmanEshel,Fisher30,Kondrashov88,Muller32} among others, have provided theoretical 
justifications that add to a long list of putative mechanisms that may account for recombination's enduring role in most 
higher species. Classic \cite{Felsenstein} and more contemporary reviews \cite{BartonAndCharlesworth,Watson} on the 
subject summarize many of these candidates. Even though the number of potential explanations is large, none of them 
has been found compelling enough to have settled the debate. Additionally, some older propositions have come under 
more scrutiny thanks to improved experimental data \cite{Kouyos,KeightleyAndOtto}, and it has even been 
suggested that the hidden value of sexual recombination might not even lie mainly in the improvement of genetic 
variability or fitness, or in its defining properties. As stated in \cite{OttoAndLenormand}:
``\ldots it is generally accepted that the long-term maintenance and ubiquity of Eukaryotic sex cannot be explained 
as an approximate consequence of the inherent properties of sex
 itself.'', a position exemplified in \cite{Baker76}, where it is suggested that recombination might serve mainly as a 
stabilizer of mitosis, and that any drawn benefit regarding genetic inheritance is circumstantial.
The plethora of proposed models ranges from simple ones that are case based 
\cite{FeldmanEshel,feldman1972selection,otto1997deleterious}, to sophisticated simulations that 
incorporate many-locus, multiple allele genotypes, dynamic recombination rates and sites \cite{Barton,liberman2008evolution,zhivotovsky1994evolution}, different levels and types of 
epistasis, mutation, complex and variable fitness landscapes, etc. \cite{Charlesworth,KeightleyAndOtto}. 
Studies typically focus on measuring the effects of recombination on average fitness, but others concentrate on 
other quantifiable benefits; \cite{Watson}, for example, reports the virtues of recombination regarding the exploration
of the fitness landscape, while in \cite{pepper2000evolution} the change over generations of the genetic linkage 
distance between epistatic units is discussed; and \cite{christiansen1998waiting} focuses on the mean time for a 
beneficial epistatic group of two alleles to appear on the same gamete with and without recombination.  
For a review on the experimental backing or counterevidence to theoretical explanations for the prevalence of 
recombination see \cite{Rice02}.

Of course, if we are to understand the benefits of recombination in the context of a mathematical model, 
a requirement is that the model itself captures the very mechanisms by which it is useful in the first place.
This then leads us to ask if the apparent inability to find an agreed universal advantage for recombination
is due to the fact that the considered models are incapable of modeling the benefits - a defect of the 
model - or, rather, that the benefits are not transparent in the analyses of the models that have been studied. 
If the models themselves are inadequate then new models with new features must be developed. On the
contrary, if the analyses themselves are at fault, one must understand why. In this paper we will start with
the hypothesis that standard population genetics models are capable of showing universal mechanisms 
by which recombination is useful. However, by restricting to a simple two-locus two-allele model we will
be able to exhaustively study the full parameter space of the model. We will show that the reason
why universal mechanisms have been difficult to identify is twofold: that the benefits are more visible in 
terms of Building Blocks (subsets of loci defined by the recombination distribution) not genotypes, as in 
standard analyses, and that the benefits of recombination are particularly associated with ``modular''
landscapes which will be discussed below. Thus, we believe, the results of this paper link two fundamental 
concepts in biology - the utility and ubiquity of recombination with the existence of modularity.

\section{Recombination - a Building Block Perspective}
\label{delta}

In this section we introduce the theoretical framework and the chief diagnostics we will use to 
examine the utility of recombination. As we are interested here in the interaction of selection and
homologous recombination we will omit mutation. We will consider the evolution\footnote{We 
will restrict attention here to a generational model with no overlap.} of a population of length 
$\ell$ haploid sequences governed by the equation \cite{stephens1}
\begin{equation}\label{dynam_eqn}
\langle P_I(t+1) \rangle = P'_I(t)-p_c\sum_m p_c(m)\Delta_I(m,t)
\end{equation}
where $\langle P_I(t+1) \rangle$ is the expected frequency of genotype $I$ at generation $t+1$. 
In the first term on the right-hand side $P'_I(t)$ is the selection probability for the 
genotype $I$. For proportional selection, which is the selection mechanism we will consider here, $P'_I(t)=(f_I/\bar f(t))P_I(t)$,
where $f_I$ is the ``survival'' fitness\footnote{By survival fitness, in the absence of factors such 
as fertility, differences in mating success etc., we mean viability, the probability to
reach reproductive age, in distinction to absolute fitness which measures the overall reproductive
success of a type.}  of genotype $I$, $\bar f(t)$ is the average population
fitness in the $t$th generation and $P_I(t)$ is the proportion of genotype $I$ in
the population. In the second term, the recombination distribution, $p_c(m)$, is 
modeled using the concept of a recombination mask $m=m_1m_2\ldots m_\ell$, which 
is such that, if $m_i=0$, the $i$th locus of the offspring is taken from the $i$th locus 
of the first parental sequence, while, if $m_i=1$ it is taken from the $i$th locus of 
the second parental sequence. Finally, $\Delta_I(m,t)$ is the Selection-weighted linkage
disequilibrium (SWLD) coefficient \cite{stephens:03} for the genotype $I$.	  
Explicitly,
\begin{equation}
\Delta_I(m,t) = (P'_I(t)-\sum_{JK}\lambda_I^{\phantom{I}JK}(m)P'_{J}(t)P'_{K}(t))
\label{dynam_eqn:p}
\end{equation}
where $\lambda_I^{\phantom{I}JK}(m)=0,\ 1$ is an indicator function that represents the 
conditional probability that the offspring genotype $I$ is formed given the
parental genotypes $J$ and $K$ and the mask $m$. For example, for two loci, $\ell=2$, 
with binary alleles, $a$ and $b$, $\lambda_{aa}^{\phantom{aa}aa,bb}(01)=0$,
while $\lambda_{aa}^{\phantom{aa}ab,ba}(01)=1$. The contribution of a particular
mask depends, as we can see, on all possible parental combinations. In this
sense, $\Delta_I(m,t)$, in the space of genotypes, is an exceedingly complicated function. 
In the case of diploids, the SWLD coefficient is equivalent to the functions $D_{i}$ of Nagylaki 
\cite{nagylaki1999convergence} and $\Theta_I$ described in \cite{burger2000mathematical}. 
For a given target genotype and mask, $\lambda_I^{\phantom{I}JK}(m)$
is a matrix on the indices $J$ and $K$ associated with the parents. For binary alleles,
for every mask there are $2^\ell\times 2^\ell$ possible combinations
of parents that need to be checked to see if they give rise to the offspring $I$.
Nevertheless, only $2^\ell$ elements of the matrix are non-zero. The question is: which ones?
Although, $\Delta_I(m,t)$, or equivalently $D_{i}$ or $\Theta_I$, gives a complete
summary of the effect of recombination in a given generation it is an exceedingly 
complicated function to analyze. However, 
the complication of $\lambda_I^{\phantom{I}JK}(m)$ in terms of genotypes is
just an indication of the fact that the latter are not a natural basis for
describing the action of recombination.

A more appropriate basis is the Building Block Basis (BBB) \cite{stephens:03,stephens_stadler}, 
wherein only the 
Building Block (BB) schemata that contribute to the formation of a genotype $I$ enter. In this case 
\footnote{Equation (\ref{dynam_eqn}) with the substitution of equation (\ref{dynam_eqn:pbb}) has a 
long history, starting with the seminal work of Hilda Geiringer \cite{geiringer44} who derived a version 
of the equation for a diploid population without selection. Versions of the equation were then rederived 
and discussed in \cite{altenberg1995schema},
who used it to discuss the performance of recombinative Genetic Algorithms using Price's theorem, 
showing that schemata were a natural consequence of recombination; and in \cite{stephens1,stephens1998effective} 
where the Building Block Hypothesis was examined and it was discussed under what circumstances 
recombination led to an increase in the effective fitness of a given genotype. 
Also, in the latter the relation to the concept of coarse graining was emphasized and discussed.}
\begin{equation}
\Delta_I(m,t) = (P'_I(t)-P'_{I_m}(t)P'_{I_{\barra{m}}}(t))
\label{dynam_eqn:pbb}
\end{equation}
where $P'_{I_m}(t)$ is the selection probability of the BB $I_m$ and
$I_{\barra{m}}$ is the complementary block such that $I_m\cup I_{\barra{m}}=I$. Both
blocks are uniquely specified by the associated recombination mask, $m=m_1m_2\ldots m_\ell$.
For instance, for three loci, $\ell = 3$, if $I=aaa$ and $m=001$ then $I_m=aa*$ and $I_{\barra{m}}=**a$,
where $*$ is the canonical ``wildcard" symbol, familiar from Evolutionary Computation, indicating that
the corresponding locus has been summed over thus leading to marginal probabilities.
Thus, the probability for the schema $x_1x_2*$ is 
$P(x_1x_2*) =\sum_{x_3=0,1} P(x_1x_2x_3)$. The selection probability for the BB
schema $I_m$ is $P'_{I_m}(t)=(f_{I_m}(t)/\barra f(t))P_{I_m}(t)$, where the fitness of $I_m$
is $f_{I_m}(t)=\sum_{I\in I_m}f_IP_I(t)/\sum_{I\in I_m}P_I(t)$ and depends on the 
actual composition of the population. It is important to emphasize that the SWLD is distinct
from the well-known linkage disequilibrium coefficient, 
$D_I(m)$, which depends only on the allele  frequencies and the crossover mask $m$,
and {\it not} on the fitness landscape. In the case of a flat fitness landscape, $\Delta_I=D_I$,
but not otherwise. In particular, a population at linkage equilibrium with $D_I=0$ does not 
necessarily satisfy $\Delta_I=0$. Selection effects generally move the system
away from the Geiringer or Robbins manifold \cite{stephens1,poli2002allele}, which is the set 
of points in the space of populations defined by $D_I=0$. In terms of BBs, 
\begin{equation}
D_I(m,t) = P_I(t)-P_{I_m}(t)P_{I_{\barra{m}}}(t)
\end{equation}
with  $P_{I_m}(t)$ and $P_{I_{\barra{m}}}(t)$ being the frequencies, not the selection probabilities, of the 
BBs $I_m$ and $I_{\barra{ m}}$. Therefore, in linkage equilibrium $D_I(m,t)=0$ implies
$P_I(t)=P_{I_m}(t)P_{I_{\barra{m}}}(t)$, i.e., the probability to find any genotype $I$ is the same as 
the product of the probabilities to find its constituent BBs. Thus, at linkage equilibrium the SWLD 
coefficient is given by
\begin{equation}
\Delta_I(m,t) = (f_I\bar f(t)-f_{I_m}(t)f_{I_{\barra{m}}}(t))\frac{P_I(t)}{{\bar f(t)}^2}
\end{equation}

Note that the structure of $\lambda_I^{\phantom{I}JK}(m)$ is particularly simple
when both $J$ and $K$ are BB schemata. For a given $I$ and $m$ one
unique BB, $I_m$, is picked out. The second BB $I_{\barra{m}}$
then enters as the complement of $I_m$ in $I$. This means that $\lambda_I^{\phantom{I}JK}(m)$
is skew diagonal on the indices $J$ and $K$, with only one non-zero element on that skew 
diagonal for a given $m$ and $I$. At a particular locus of the offspring, the associated allele is taken 
from the first or second parent according to the value of $m_i$. If it is taken from the first parent,
then the corresponding allele in the second parent is immaterial. As seen above, this fact is
represented by the normal schema wildcard symbol $*$. It is important to emphasize that the 
BBs form an alternative basis to that of the genotypes. This means that genetic dynamics
can not only potentially be described without any reference to genotypes but also that with the 
dynamics of the BBs the dynamics of any and all genotypes can be derived. 
For instance, for two loci with binary alleles, $a$ and $b$, the possible genotypes are $bb$, 
$ba$, $ab$ and $aa$. The corresponding BBs are $aa$, $a*$, $*a$ and $**$, where 
we arbitrarily chose the genotype $aa$ as the type around which to develop the BBB. The 
relationship between the two bases is given by 
\begin{equation}
\left(\begin{array}{c}
P_{**}\\
P_{*a}\\
P_{a*}\\
P_{aa}
\end{array}\right)
=\left(\begin{array}{cccc}
1 & 1 & 1 & 1\\
0 & 1 & 0 & 1\\
0 & 0 & 1 & 1\\
0 & 0 & 0 & 1
\end{array}\right)
\left(\begin{array}{c}
P_{bb}\\
P_{ba}\\
P_{ab}\\
P_{aa}
\end{array}\right)
\label{coordtrans}
\end{equation}
where
\begin{equation}
\Lambda_{BB}=\left( \begin{array}{cccc}
1 & 1 & 1 & 1\\
0 & 1 & 0 & 1\\
0 & 0 & 1 & 1\\
0 & 0 & 0 & 1
\end{array}\right)
\end{equation}
is the coordinate transformation matrix that transforms from one basis to another. As bases, the genotype
and BBB have equivalent dynamics. However, the dynamics of recombination is fundamentally simpler in 
the BBB due to the immense simplification of $\lambda_I^{\phantom{I}JK}(m)$ in the latter. In other words,
just as Walsh/Fourier modes \cite{goldberg1,vose_wright:walshII,Weinberger:91b,wright00:_exact_schem_theor} 
are the natural basis for describing mutation, so BB schemata are the natural basis for 
describing homologous recombination. They are the natural effective degrees of freedom of
any genetic system with recombination.

From Equation (\ref{dynam_eqn}) for the time evolution of the probability distribution for 
the system, we may derive the time evolution of any derived quantity, such as the average population
fitness, which is given by
\begin{equation}
\langle \bar f(t+1) \rangle = \sum_I \frac{f^2_I}{\bar f(t)}P_I(t)-p_c\sum_mp_c(m)\sum_If_I\Delta_I(m,t)
\label{fitness}
\end{equation}
\subsection{Why Recombination?}
\label{why}

As mentioned in the introduction, a great amount of work has been done on trying to understand
why recombination is ubiquitous. Here, rather than trying to understand the potential 
benefits of homologous recombination at the most general phenomenological or conceptual level, 
we will restrict attention to what we may deduce purely from its mathematical representation in 
equation (\ref{dynam_eqn}). Of course, it may be that the benefits of recombination are not
manifest in this model. However, given that the model is the generally accepted framework for
classical population genetics it behooves us to at least use it as a starting point.
Further, we will analyze the model concentrating on two simple metrics for measuring the 
benefits of recombination, asking: i) under what circumstances can recombination lead to the 
generation of a higher frequency of a fit offspring than would be the case with only
selection? and, relatedly, ii) under what circumstances can recombination lead to 
a larger increase in the average population fitness relative to selection only? From
equations (\ref{dynam_eqn}) and (\ref{fitness}) we see that it is the SWLD coefficient 
that quantifies the effect in both cases. 

From equation (\ref{dynam_eqn}), we can see that if $\Delta_I(m) < 0$ then
recombination leads, on average, to a higher frequency of the genotype $I$
than in its absence. In other words, in this circumstance, recombination
is giving you more of $I$ than you would have otherwise. On the contrary,
if $\Delta_I(m) > 0$ then the converse is true, recombination provides
less of the genotype of interest than would be the case in its absence. With this is mind,
as mentioned, we will consider two complementary metrics to evaluate the utility of 
recombination in time: the change in number of optimal genotypes from one generation 
to the next and the change in average population fitness. In the infinite population limit, 
the former is given by
\begin{equation}
\Delta_{P_I}(t)=P_I(t+1)-P_I(t)=(P'_I(t)-P_I(t))-p_c\sum_m p_c(m)\Delta_I(m,t)
\end{equation}
For fitness-proportional selection, 
\begin{equation}
\Delta_{ P_I}(t)= \left(\frac{f_I}{\bar f(t)}-1\right)P_I(t)-p_c\sum_m p_c(m)\Delta_I(m,t)
\label{deltapchange}
\end{equation}
The first term on the right-hand side is the increase in the number of optimal genotypes 
due to the effect of selection only and the second term the contribution due to recombination.
Now passing to the average population fitness, we can consider two reference points for
measuring the effect of recombination relative to selection. The first is to consider  
in the infinite population limit
\begin{equation}\label{DeltaEfBar}
\delta_{ \bar f}(t)=(\barra{f^2}-{\bar f}^2)-p_c\sum_m p_c(m)\sum_I f_I\Delta_I(m,t)
\end{equation}
where, once again, the first term on the right-hand side is the contribution from selection only,
and corresponds to Fisher's Fundamental Theorem, while the second term is the contribution 
from recombination. In both $\Delta_I$ and $\delta_{ \bar f}(t)$, we are considering metrics that 
measure the relative contribution of recombination generation by generation, {\it not} the cumulative effect of
recombination versus selection. As a measure of the latter we consider
\begin{eqnarray}
 \Delta_{\bar f}(t)&=&\overline{f}_{\mbox{\tiny r+s}}(t)-\overline{f}_{\mbox{\tiny s}}(t),\\
&=&\frac{\barra{f^2}_{s+r}(t-1)}{{\bar f}_{s+r}(t-1)}-p_c\sum_m p_c(m)\sum_I f_I\Delta_I(m,t)-\frac{\barra{f_{s}^2}(t-1)}{{\bar f}_{s}(t-1)}
\label{DELTAf}
\end{eqnarray}
Thus, if $\Delta_{\bar f}(t)$ is positive then the average fitness of the population evolving in the presence
of recombination and selection (s+r) is higher than that of the same population evolving in the presence of 
selection only (s). 

For both, generation by generation metrics the qualitative contribution of recombination is purely 
controlled by the sign of $\Delta_I(m)$. For increasing the frequency of a fit genotype $I$ relative to the case of selection only,
we see that this will be the case, passing from generation $t$ to generation $t+1$, 
if and only if $\Delta_I(m,t)<0$, with the sign and magnitude of $\Delta_I(m,t)$ fixed 
completely by the fitness landscape and the actual population. So, whether recombination 
is beneficial or not passing from one generation to another, in this sense, is equally fixed by the
fitness landscape and the actual population. Similarly, the increase in the average population
fitness from one generation to the next, relative to selection only, is controlled by the fitness 
weighted average of $\Delta_I(m,t)$ and, hence, once again, by the fitness landscape and the current
population. However, in the case of the cumulative measure we see that the potential contribution of recombination
is more subtle as besides the explicit term $p_c\sum_m p_c(m)\sum_I f_I\Delta_I(m,t)$ there is also the effect
of the difference between $\barra{f^2}_{s+r}(t-1)/{\bar f}_{s+r}(t-1)$ and $\barra{f^2}_{s}(t-1)/{\bar f}_{s}(t-1)$
which depends implicitly on $p_c$.
%
%


So, once again, we are led to ask first: When is $\Delta_I(m,t)<0$? The answer is when 
$P'_I(t)< P'_{I_m}(t)P'_{I_{\barra{m}}}(t)$, i.e., the probability to select the genotype $I$ is less than the
probability to select its component BB schemata, where the action of recombination is modeled to 
be such that the blocks are selected independently. There are several distinct regimes in which 
$\Delta_I(m,t)<0$, which we will explore further and which categorize the different conditions under
which homologous recombination can be deemed useful. First, there is the regime in which $P_I(t)=0$, i.e., 
the genotype $I$ is non-existent, or at a very small frequency, in the actual population. 
In this case $\Delta_I(m,t)<0$ directly and 
then, remembering that we are neglecting the effects of mutation, recombination is the only 
mechanism by which the genotype $I$ can be generated. This regime emphasizes the search 
property of recombination, independent of the fitness landscape. 

In general though, as emphasized, the effects of recombination depend on the fitness landscape.
Taking the classic Muller's ratchet argument as a reason why recombination exists it has been 
shown that modifier genes that lead to higher recombination rates could increase in the presence 
of negative {\it multiplicative} epistasis \cite{Charlesworth,feldman1970quasilinkage,kondrashov1984deleterious,charlesworth1993effect}.
However, if the epsitasis was too great the effect disappeared. Thus, in the parameter space for the 
landscape the advantage for recombination only appeared in a smal region and therefore could not be
offered as a generic explanation for the ubiquity of recombination and sex. In other work, 
\cite{stephens2007just,stephens_cervantes} have provided evidence that recombination is particularly
beneficial in an additive landscape with zero {\it additive} epistasis and very detrimental in a
landscape with high positive {\it additive} epistasis. A simple way to see this is to eliminate any bias 
that comes from a particular choice of initial population and assume equal proportions for all genotypes.
In this situation, it can be shown that 
$\Delta_l(m,t) = (f_I\bar f(t)-f_{I_m}(t)f_{I_{\barra{m}}}(t))/2^\ell{\bar f(t)}^2 < 0$ for {\it any}
$m$ that does not cut an epistatic link between loci. For instance, for a genotype $I_1I_2\ldots I_\ell$,
if $f_I=\sum_{I_i} f_{I_i}$, i.e., the landscape is additive, then $\Delta_l(m,t) < 0$ for any $m$. 
This result is also valid when the $I_i$
correspond to multiple loci when recombination does not cut any epistatic link between the loci. This
is the case for a modular landscape, where loci divide up into disjoint sets with epistasis 
between the loci in a set but not between sets.  The benefit of recombination in this case is that it efficiently
increases the number of fit non-epistatically linked BBs in an offspring genotype relative to 
the numbers present in the parental types. On the contrary, for a highly additively epistatic 
fitness landscape, such as ``needle-in-a-haystack" (NIAH)\footnote{This landscape corresponds to one optimal 
genotype with fitness $f_n$, while other types have equal fitness, $f_h$. It has been used extensively in
molecular evolution in the context of the Eigen model \cite{eigen}, where the dynamics is 
naturally understood in terms of quasi-species.} one can show that $\Delta_l(m,t) > 0$ for all $m$.
As is well known, for a multiplicative landscape, $\Delta_l(m,t) = 0$.

One may argue, of course, that proving that $\Delta_I(m,t) < 0$ over one generation for a particular 
choice of population and in particular fitness landscapes does not correspond to a ``universal'' 
mechanism for explaining the benefits of recombination. That is
why in this paper we consider the general situation of an arbitrary fitness landscape and an arbitrary 
population, as well as considering multiple generations. To consider such generality, however, the price we 
must pay is to restrict to a small number of loci.

So, we would argue that two significant, and potentially related, regimes in which 
recombination is beneficial are: i) the {\it search} regime, where recombination searches for fit 
genotypes that presently either do not exist or are at very low frequency in the population; 
and ii)  the {\it modular} regime, with either weak positive or negative additive epistasis, 
where recombination allows for the juxtaposition of distinct 
fit modules in different parental types into an even fitter offspring. Of course, in the {\it search} 
regime the question arises as to whether recombination is more efficient than mutation. This 
will depend on the Hamming or edit distance between parents and offspring.
An example, that we will not consider in more detail, that exhibits the benefits of recombination over
mutation in generating innovation, is the development of antibiotic resistance in bacteria through
horizontal gene transfer. Generically, it will be the case that the Hamming or edit distance 
between the original parental sequences, say bacterium and virus, and the offspring sequence,
bacterium with viral gene, will be potentially large. In other words, the difference
between the initial and final sequences is not a single-nucleotide, or even a small number of them.
In this sense, recombination-like\footnote{By ``recombination-like'' we mean any genomic change where
one or more sub-sequences in one or more parental sequences are transferred to an offspring sequence.
This is termed ``generalized recombination'' in \cite{poli_stephens_genrecomb} and comprehends 
unequal crossing over, transposition, translocation and related operations, as well as homologous recombination.} 
events are the only way to generate innovation that is associated
with large genomic changes, ``large'' meaning that the Hamming or edit distance between parental
and offspring sequences is large.    
\section{Modularity and Fitness Landscapes}
\label{module}

Before considering our explicit model we wish to discuss the concept of modularity in terms of the 
fitness landscape. For simplicity, we restrict to binary alleles $x_{i}=0,\ 1$, where $i$ refers to
the locus. We will consider two representations of the fitness function, a direct one where we use 
the $f_x=f_{x_{1}x_{2}\ldots x_{\ell}}$ directly and another one where the fitness function can be 
written as an expansion of the form
\begin{eqnarray}
f_{x}&=&F^{(0)}
+\sum_{i_1=1}^\ell F_{i_1}^{(1)}x_{i_1}+\sum_{i_1=1}^{\ell-1}\sum_{i_2=i_1+1}^{\ell}
F_{i_1i_2}^{(2)}x_{i_1}x_{i_2}\label{land_exp}\\
&+&\sum_{i_1=1}^{\ell-2}\sum_{i_2=i_1+1}^{\ell-1}\sum_{i_3=i_2+1}^{\ell}
F_{i_1i_2i_3}^{(3)}x_{i_1}x_{i_2}x_{i_3}
+\ldots+ F_{i_1i_2\ldots i_\ell}^{(\ell)}x_{i_1}x_{i_2}\ldots x_{i_\ell}
\nonumber
\end{eqnarray}
where $F^{(n)}_{i_1i_2\ldots i_n}$ represents an epistatic interaction between $n$ alleles
located at loci $i_1\ ,i_2\ ,\ldots\ ,i_n$ and $x_{i_n}=0,\ 1$. The advantage of this latter representation 
is that the degree of epistasis between different loci and alleles can be simply deduced. 

Any landscape that contains only Fourier components
of $O(n)$ is said to be an elementary landscape of order $n$. 
For instance, a completely additive landscape has a fitness function of the form 
\[
f_{x}=\sum_{i=1}^\ell F_ix_i
\]
and is therefore an elementary landscape of order one, as all Fourier components
other than order one are zero. This is a consequence of the fact
that there are no epistatic interactions between loci. Similarly,
a multiplicative landscape, where 
\[
f_{x}=F_{i_1i_2\ldots
  i_\ell}^{(\ell)}x_{i_1}x_{i_2}\ldots x_{i_\ell}
\]
is an elementary landscape of order $\ell$, as all Fourier components
other than order $\ell$ are zero, there being epistatic
interactions of order $\ell$ between the loci but no others. 
Other landscapes will be intermediate between these extremes. 
Once again, we emphasize here that we are measuring epistasis relative to the additive 
limit not the multiplicative one as has been the norm in most papers on recombination
and population genetics.

A particularly interesting class of landscapes in terms of their relevance for recombination 
are those of ``modular'' type, where the loci 
of a genotype partition into $\ell_m$ disjoint subsets\footnote{Intuitively these modules will be formed 
by contiguous loci such as is natural for an exon or gene.}, modules, $s_1,s_2,\ldots s_{\ell_m}$.
We will consider two complementary notions of modularity here, one where 
the landscape can be decomposed as the sum of the individual fitnesses of these disjoint subsets,
and one where the fitness is associated with a Boolean "OR" function on the alleles of the modules.
In the first case the fitness of a genotype is given by
\begin{equation}
f_{x} = \sum_{s_i=1}^{\ell_m} f_{s_i}
\end{equation}
the sum of the fitnesses of its constituent modules.
This modularity will obviously leave an imprint in the expansion (\ref{land_exp}). For instance,
if each module consists of $\ell_m$ loci and there is no epistasis between the modules then in 
(\ref{land_exp}) we will have $F_{i_1i_2\ldots  i_n}^{(n)}=0$ for $n>\ell_m$. In the second case,
our notion of modularity is associated with the idea of genetic redundancy, whereby the fitness
of a genotype is similar in the presence of different copy numbers of a given gene. The extreme
limit of this is when the landscape is associated with an ``OR" function, so that the fitness of a type
is the same whether there is one or multiple copies of a gene. The intuition of a module in this
context is that in the presence of redundancy with multiple copy number one, or maybe more, 
genes can be removed or mutated without affecting the fitness of the type. Thus, a gene acts
as a module as it can be changed independently without affecting the fitness of the type. As 
we will see, this corresponds to a system with a maximal degree of negative epsitasis. 

As mentioned previously, a full analysis for $\ell$ loci with arbitrary landscape and population
is prohibitively difficult, so here we will focus on the case of two loci, as in this case
we can study in the context of an exactly solvable model the different regimes under which 
recombination can be beneficial. So, restricting ourselves to the case of two loci, $\ell=2$, we have
\begin{eqnarray}
f_{x_{1}x_{2}}=F^{(0)}
+\sum_{i_1=1}^2 F_{i_1}^{(1)}x_{i_1}+F_{12}^{(2)}x_{1}x_{2}
\label{stat_phys_rep_fitness}
\end{eqnarray}
For an additive (modular) landscape $F^{(2)}_{12}=0$. For a multiplicative landscape 
$F^{(0)}F_{12}^{(2)}=F_{1}^{(1)}F_{2}^{(1)}$. 
For a redundant (modular) landscape $F^{(2)}_{ij}=-F^{(1)}_{i}=-F^{(1)}_{j}$ which, as mentioned,
can be understood in terms of a Boolean "OR", fitness being the same if either one or both alleles
are optimal. For a NIAH landscape $F_{1}^{(1)}=F_{2}^{(1)}=0$ which, in contrast to the redundant
landscape corresponds to a Boolean "AND" as fitness is only different if both alleles are optimal.
\section{Recombination in an exact two-locus model}
\subsection{Analytic results}

Clearly, trying to characterize the efficacy of recombination quantitatively, and in detail, is 
prohibitively complicated. As we saw in section \ref{delta}, however, within the confines of the model 
we are considering, in a given generation, it can be characterized using only one fundamental function: the SWLD 
coefficient. The SWLD coefficient, though, depends not only on the recombination distribution, 
but also on the fitness landscape and the current state of the population. In other words it is 
a function of a large number of parameters. To circumvent this problem we consider the case of two loci
and calculate the SWLD coefficient as a function of the fitness landscape and the population. Note that by
two loci here we do not necessarily imply that they represent ``genes". They may represent any two 
structural units, such as exons, introns or other motifs, or nucleotides themselves, that
can be separated or recombined by crossover and which can be characterized, as an approximation, 
by a fitness landscape that is independent of the rest of the genome.  

\begin{figure}[hbt]
\centerline{\psfig{file=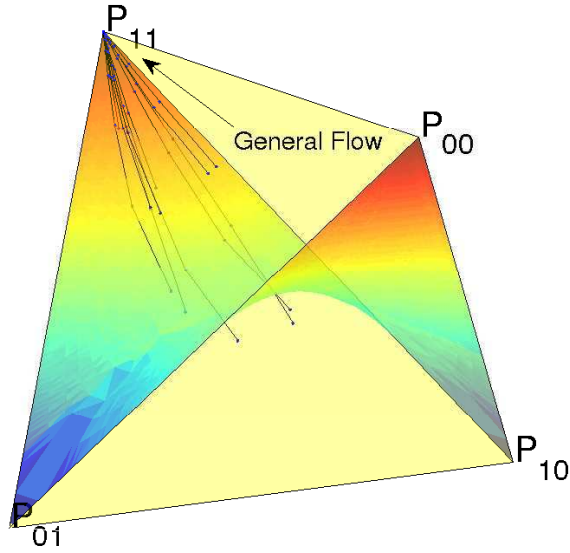,width=11.5cm}}
\caption{Geiringer manifold (colored) and some trajectories for some random initial populations. The system's 
convergence to dominance of the optimal genotype is indicated by the arrow.}
    \label{GeiringerAndTrajectories}
\end{figure}

For two loci all genotypes can be characterized by a multi-index $I=ij$, with 
$i,j\in\{ 0,1,\ldots,\mathcal C\}$, where $\mathcal C+1$ is the cardinality of the 
alphabet that labels the loci, or alleles in the case of genes. For $\ell=2$, there is only one 
non-trivial mask\footnote{The masks $m=00$ and $11$ correspond to cloning, where both offspring
loci come from a single parent.} $m=01$, and its conjugate, that lead to the BBs $i*$ and $*j$. 
The sum over masks in the general expression for the SWLD coefficient is thus reduced  
to only one term:%
\begin{equation}%
\Delta_{ij}=P'_{ij}-P'_{i*}P'_{*j}=P'_{ij}-(P'_{ii}+P'_{i\Not{i}})(P'_{jj}+P'_{\Not{j}j}),%
\end{equation}%
Direct evaluation shows that%
\begin{equation}%
 \Delta_{ij}=\Delta_{\Not{i}\Not{j}}=-\Delta_{i\Not{i}}=-\Delta_{\Not{j}j},%
\label{DeltaEquality}%
\end{equation}%
and thus the evolution equations in the two-allele, two-locus problem are:%
\begin{equation}
 P_{ij}(t+1)=P_{ij}'(t)- p_c \Delta_{ij}\\
\end{equation}
The whole state of this system can be characterized by 3 $(=4-1)$ frequencies that are naturally 
represented in a three dimensional simplex. Figure \ref{GeiringerAndTrajectories} 
shows typical population trajectories in the two-locus, two-allele system for a generic 
landscape, with $x=11$ arbitrarily taken as the optimum genotype and several 
different initial population ratios.

As far as the fitness landscape is concerned the general parametrized two-locus two allele landscape is
\begin{equation}
f=a+b_1x_1 +b_2x_2+cx_1x_2
\end{equation}
where $c$ is the measure of the {\it additive} epistasis between the two loci. We take
the genotype $I=00$ as the wild type, the genotypes $I=01$ and $10$ as single mutants and $I=11$ 
as a double mutant which is the optimal genotype. 
There are just three main landscape categories for the two-bit, two-locus model:
\begin{enumerate}
 \item The wild type and the double mutant are the anti-optimum and optimum respectively.
\item One of the single mutants (10 or 01) is the antioptimum.
\item The two lowest fitness phenotypes are the single mutants.
\end{enumerate}
Any other case can be brought to one of the previous by a relabeling that doesn't affect the dynamics.
In the first two landscape types, a generic population will always eventually evolve towards the global optimum. 
In the third type, the population may converge to the optimum or the suboptimal wild type
$00$ depending on the initial population and the recombination probability.\footnote{The latter
two landscape categories are known as {\it deceptive} landscapes of Type I and Type II respectively 
in the Genetic Algorithm literature\cite{goldberg1}. It has been proved \cite{takahashi1998convergence} that Type I 
systems always converge to the global optimum whereas Type II systems converge to the optimum
or double mutant depending on the population and recombination probability. }

From Equations (\ref{deltapchange}) and 
(\ref{DELTAf}) we have 

\begin{equation}
\Delta_{P_{ij}}(t)=\left(\frac{f_{ij}}{\bar f(t)}-1\right) - p_c\Delta_{ij}(t)
\label{metric1twobit}
\end{equation}
\begin{equation}
\Delta_{\bar f}(t)=\frac{\barra{f^2}_{s+r}(t-1)}{{\bar f}_{s+r}(t-1)}-\frac{\barra{f_{s}^2}(t-1)}{{\bar f}_{s}(t-1)}
 - p_c\sum_{ij} f_{ij}\Delta_{ij}(t)
\label{metric2twobit}
\end{equation}
For the optimal genotype
\begin{equation}
\Delta_{11}(t) = \frac{1}{{\bar f}^2(t)}(a(a+b_1+b_2+c)P_{11}(t)P_{00}(t)-(a+b_1)(a+b_2)P_{01}(t)P_{10}(t))
\label{twobitdeltafull}
\end{equation}
As mentioned, the sign of $\Delta_{ij}$ determines the qualitative effect of recombination in a given generation.
To develop some intuition for how the characteristics of the landscape affect our metrics we will set for the moment
$P_{ij}(t)=1/4$, i.e., a homogeneous population with no initial bias for one genotype versus another. 
As the parameter $a$ just sets the scale for the landscape we can without loss of generality for fitness proportional
selection set $a=1$. We will also set $b_1=b_2=b$ so that both single mutants have the same fitness. In this case,
\begin{equation}
\Delta_{11}(t) = \frac{(c-b^2)}{(1+b+c/4)^2}
\label{deltatwobithomo}
\end{equation}

For a multiplicative landscape $c=b^2$ and $\Delta_{11}=0$, as is well known. For an additive landscape $c=0$
and therefore $\Delta_{11}(t)=-b^2/(1+b)<0$. In this case recombination leads to a higher frequency of the optimal
genotype in the next generation than selection alone. For a deceptive landscape, $b<0$, but $c>-2b$ and so $\Delta_{11}(t)>0$
and recombination in this region of the parameter space leads to a lower frequency of the optimal genotype in the 
next generation. In terms of BBs, for deceptive landscapes, the marginal fitnesses are such that $f_{1*}<f_{0*}$ and 
$f_{*1}<f_{*0}$, and so the reason why recombination is unfavourable is that the necessary mutant alleles for 
constructing the optimal genotype are deleterious relative to the corresponding alleles of the genotype $00$.  
For additive epistasis, such that $c> b^2$, we have $\Delta_{11}(t)>0$ and recombination once 
again leads to a lower frequency of the optimal genotype in the 
next generation than selection alone. Generally, if we take $c-b^2<0$ as signifying negative multiplicative epistasis then 
we see that in such landscapes recombination has a positive effect in terms of our $\Delta$ metric and on the contrary
for positive multiplicative epistasis. Note that the additive limit $c=0$ corresponds to negative multiplicative epistasis.
Interestingly, equation (\ref{deltatwobithomo}) shows that the greatest benefit from recombination, i.e., the minimum value 
of $\Delta_{11}$ is associated with landscapes with negative {\it additive} epistasis, i.e., $c<0$. Maximum negative
epistasis is given by the minimum value of $c$, $c=-b$. In this case $\Delta_{11}(t)=-b(1+b)/(1+3b/4)^2$.

Why would this maximum negative epistasis be associated with the utility of recombination, at least in terms of metric 
(\ref{deltapchange})? Examining equation (\ref{twobitdeltafull}) we see that the first term, proportional to $P_{11}(t)P_{00}(t)$,
corresponds to elimination of the optimal genotype $11$ by recombining it with the suboptimal genotype $00$, whereas  
the term proportional to $P_{01}(t)P_{10}(t)$ corresponds to construction of $11$ via recombination of the 
single mutants $10$ and $01$. It is the competition between these two effects that measures the benefits of recombination
in terms of (\ref{deltapchange}). Additive landscapes with $c=0$ reduce the impact of
destruction without compromising the positive effect of reconstruction. Negative epistasis, on the other hand, does not 
affect the construction of the optimal genotype by recombining the single mutants, but it does minimize the effect
of destruction of the optimal genotype. The maximal effect is when $c=-b$ and corresponds to a Boolean "OR" 
landscape where $f_{01}=f_{10}=f_{11}> f_{00}$. This is the situation where there is genetic redundancy, as the 
fitness of the optimal phenotype requires the presence of only one optimal allele not both. 
At this naive level we also see that the benefit of recombination is not 
restricted to small negative multiplicative epistasis but, rather, the larger the additive negative epistasis the larger 
the benefit conferred by it. 

In terms of the metric (\ref{DeltaEfBar}) the contribution from recombination is given by
\begin{equation}
\sum_{ij} f_{ij}\Delta_{ij}(t)=c\Delta_{11}(t)=\frac{c(c-b^2)}{(1+b+c/4)^2}
\label{metric2twobitcontrib}
\end{equation}
For this term to give a positive contribution to the average population fitness we require $c(c-b^2)<0$. For $c>0$
this requires $c<b^2$, which we will term weak positive additive epistasis. On the other hand, for $c<0$, $c(c-b^2)>0$ and 
recombination apparently leads to a decrease in the average population fitness, while in the additive limit, $c=0$,
there is no change. Together, a one generation analysis of our two metrics would indicate that there are benefits
to recombination from both of them only for weakly positively additively epistatic landscapes such that $c>0$ and $c<b^2$.
We will characterize these landscapes as being ``modular'', i.e., quasi-additive.
It is important however, to go beyond a single generation, and for that we will consider metric (\ref{DELTAf}) in
section \ref{ParameterSweep}.

\subsubsection{Muller's Ratchet.}
\label{MullersRatchet}

Muller's ratchet \cite{Muller32}\footnote{A good, although somewhat dated, review of the different potential
mechanisms, and in particular Muller's ratchet, by which recombination can be beneficial can be found 
in \cite{Felsenstein}.}, and variations thereof, have been frequently invoked in considerations of the 
potential benefits of recombination. Essentially, the argument is that recombination increases the 
evolvability of a population by allowing beneficial mutations on different genomes to be recombined
into one more efficiently than the process of generating a double mutation. Similarly, deleterious 
mutations can be eliminated more efficiently from a population by having them recombined into
a single genome, thus allowing selection to eliminate them more efficiently. We will consider these 
arguments in the context of our two locus system. 

There are two regimes of interest related to Muller's ratchet, one is that advantageous mutations
appear in a population and the second that deleterious mutations appear. The question is: How does
recombination affect the dynamics of these mutants? Considering the first case, if
we consider the population to be such that the fit double mutant is absent, i.e., 
$P_{11}(t)=0$,\footnote{In this case there is an initial linkage disequilibrium, i.e.,\\ 
$(P_{11}(t) P_{00}(t)- P_{10}(t) P_{01}(t))\neq 0$. } then 
$\Delta_{11}=(P'_{11}(t)P'_{00}(t) -P'_{10}(t) P'_{01}(t))=-P'_{10}(t)P'_{01}(t)< 0$. So
\begin{equation}
\Delta_{P_{11}}(t)=\left(\frac{f_{11}}{\bar f(t)}-1\right)P_{11}(t) +p_cP'_{10}(t)P'_{01}(t)
\label{deltaoptimummullerad}
\end{equation}
\begin{equation}
\delta_{\bar f}(t)=(\bar{f^2} - {\bar f}^2) + p_c(f_{11}-f_{01}-f_{10}+f_{00})P'_{10}(t)P'_{01}(t).
\label{deltafbarmullerad}
\end{equation}
From Equation (\ref{deltaoptimummullerad}) we see that the number of fit double mutants
increases from generation $t$ to generation $t+1$ due to the effect of recombination relative 
to selection only dynamics. This is, in fact, independent of the fitness landscape, being associated
with the {\it search} regime of recombination alluded to in section \ref{why}. In contrast, in Equation
(\ref{deltafbarmullerad}), we see that the average population fitness will increase in the presence of
recombination if and only if $F^{(2)}=(f_{11}-f_{01}-f_{10}+f_{00}) > 0$, which is a direct measure of 
the degree of {\it additive} epistasis between the two loci. As noted, for a purely additive landscape,
$F^{(2)} = 0$ and so recombination is neutral in this setting. For the other genotypes we have the 
fraction of wild types increases due to the effect of recombination, while the frequency of single mutants
decreases. What happens in the case where $P_{11}(t)\neq 0 $ will be considered in section 
\ref{ParameterSweep} as the benefit from recombination then depends on the actual population 
as well as the landscape. 

Turning now to the case of deleterious mutants: in this case we take the wild type to be the genotype 
$11$ and the types $01$ and $10$ to be deleterious single mutants and $00$ to be an even 
more deleterious double mutant. In this case, just as for beneficial mutants, 
$\Delta_{11}=(P'_{11}(t)P'_{00}(t) -P'_{10}(t) P'_{01}(t))=-P'_{10}(t)P'_{01}(t)< 0$ and hence the proportion
of optimal wild types $11$ increases. In terms of average population fitness, the increase from generation
$t$ to $t+1$ is given by Equation (\ref{deltafbarmullerad}). In other words the change in average population
fitness per generation for the case of beneficial versus deleterious mutations is identical if we are 
considering the same fitness landscape.

\subsubsection{Asymptotic behavior of $\Delta$}
\label{Asymptotic} 

Before going on to consider the full numerical solution of the two-locus model we will consider 
what can be said analytically about the asymptotic behavior of the system.  Although there are 7 parameters 
that control the dynamics, the asymptotic behavior can be most naturally written in terms of just 
two parameters
\begin{equation}
C(t)\equiv \frac{P_{11}P_{00}}{P_{10}P_{01}},
\end{equation}
where, for brevity, we use $P_{ij}$ for $P_{ij}(t)$, and
\begin{equation}
A\equiv \frac{f_{10}f_{01}}{f_{11}f_{00}}.
\end{equation}
The one generation evolution equation for $C(t)$ is
\begin{eqnarray}
 C(t+1)&=& \frac{P_{11}(t+1)P_{00}(t+1)}{P_{10}(t+1)P_{01}(t+1)}=\frac{(P'_{11}-p_c\Delta)(P'_{00}-p_c\Delta)}{(P'_{10}+p_c\Delta)(P'_{01}+p_c\Delta)}\nonumber \\
&=&\frac{\frac{P'_{11}P'_{00}}{\Delta}-p_c(P'_{11}+P'_{00})+p_c^2\Delta}{\frac{P'_{10}P'_{01}}{\Delta}+p_c(P'_{10}+P'_{01})+p_c^2\Delta}
\label{CEvolution}
\end{eqnarray}

Without loss of generality we again choose $I=11$ to be the optimal genotype. The evolution of the genotype  
frequencies, $P_{ij}$, as given by equation (\ref{dynam_eqn}), ensures the eventual dominance of 
one of the genotypes\footnote{Karlin, see for example \cite{karlin1975general} section vii, has shown 
that there are no stable polymorphisms in the model type considered in this paper.}. The first part of 
this derivation is analogous to section 3 in \cite{feldman1970quasilinkage}. We suppose {\it a priori} that the limit
\begin{equation}
 C_\infty\equiv\lim_{t\rightarrow \infty} C(t)
 \end{equation}
exists, which in turn implies that 
\begin{equation}
 \lim_{t\rightarrow \infty} \frac{P'_{11}P'_{00}}{\Delta}=\frac{C_\infty}{C_\infty-A}
 \end{equation}
and 
\begin{equation}
 \lim_{t\rightarrow \infty} \frac{P'_{10}P'_{01}}{\Delta}=\frac{A}{C_\infty-A}
 \end{equation}

 With these elements in hand we can calculate the putative limit of equation (\ref{CEvolution}) to find:
\begin{equation}
 C_\infty=\frac{\frac{C_\infty}{C_\infty-A}-p_c}{\frac{A}{C_\infty-A}},
 \end{equation}
Solving this last equation for $C_\infty$ we obtain:
\begin{equation}
 C_\infty=\frac{p_c A}{p_c+A-1},
 \end{equation}
Finally, since $\Delta=P'_{11}P'_{00}-P'_{10}P'_{01}$ and $C'=\frac{P'_{11}P'_{00}}{P'_{10}P'_{01}}$, we note that the negativity
 of $\Delta$ is equivalent to the condition
\begin{equation}\label{AMoreThanOne}
 C'_\infty\equiv \lim_{t\rightarrow \infty} \frac{P'_{11}P'_{00}}{P'_{10}P'_{01}}\equiv\frac{C_\infty}{A}=\frac{p_c }{p_c+A-1}<1,
 \end{equation}
which reduces to $A>1$ for $p_c\neq 0$. So, we can see that the asymptotic benefit of recombination 
in terms of increasing the fraction of optimal genotypes relative to selection only, is determined by 
only 2 parameters - $A$ and $p_c$ and is independent of the initial population.

With this formula in hand, we can easily map any fitness landscape to a range of values for $A$ 
and thus determine if recombination will be asymptotically favorable for that particular landscape.
we have
\begin{equation}
A=\frac{(a+b_1)(a+b_2)}{a(a+b_1+b_2+c)},
\end{equation}

To simplify further the visualization of the 
asymptotic behavior, we again assume that $b\equiv b_1=b_2$, i.e., that the two mutants have the 
same fitness. As eventually the optimal genotype dominates for non-deceptive landscapes, recombination is
asymptotically neutral. However, how $\Delta$ approaches zero depends on $A$. 
Small values of values of $b$ and $c$ correspond to a more neutral
fitness landscape, where selection effects are small. For an additive 
landscape $A>1$ and so recombination is asymptotically 
beneficial in that $\Delta$ tends to zero from negative values. 
Small values of $b$ relative to $c>0$ correspond to highly positively additively epistatic landscapes and in this case
$A<1$ and recombination is asymptotically disadvantageous in that $\Delta$ approaches zero 
from positive values. The multiplicative landscape with $c=b^2$ and, 
hence, $A=1$, separates the two classes of behavior. 
The dependence of the parameter $A$ (=$\frac{f_{01}f_{10}}{f_{00}f_{11}})$ 
as a function of $b$ and $c$ is shown in the next graph:
Values of $A$ greater than $1$ mean that the iterates must eventually reach negative values 
of $\Delta$. The sign of $\Delta$ is then conserved, although the magnitude approaches zero as the 
system reaches linkage equilibrium associated with a population dominated by the optimal genotype. 
The opposite happens when $A<1$. Note that the locus defined by the intersection of the surfaces $A(b,c)$
and $A=0$ is given by $b^2=c$ and corresponds to the case of multiplicative landscapes. 

\ImagehereFull{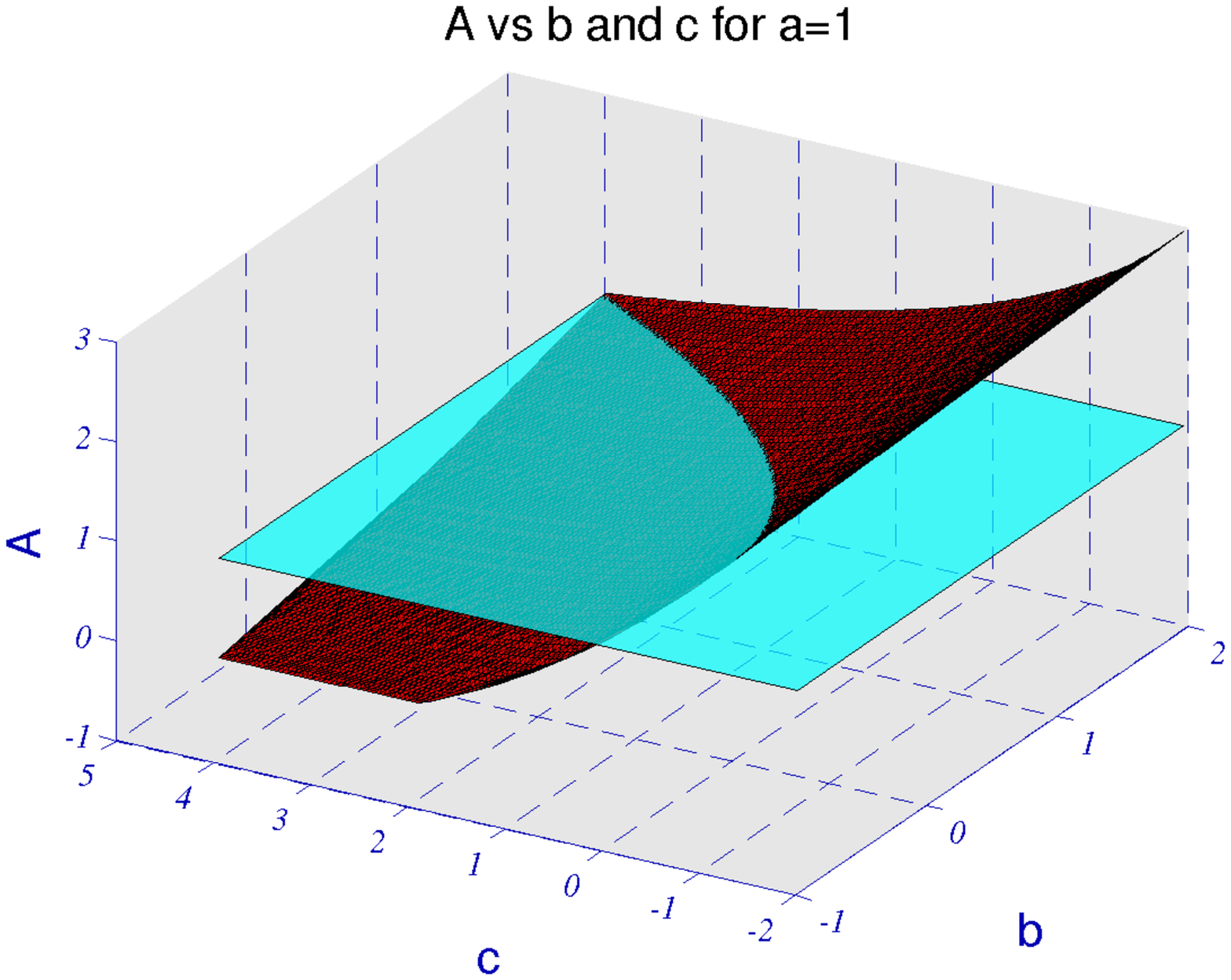}{11.5cm}
{ $A(b,c)=\frac{f_{01}f_{10}}{f_{00}f_{11}}$ for $a=1$. The solid plane, $A=1$, separates those fitness 
landscapes that according to Eq.\ref{AMoreThanOne} will eventually benefit from recombination from 
those that don't. }{FiguraUno}

%
%
\section{Exact Numerical Results}
\label{ParameterSweep}

Turning now to the non-asymptotic behavior, we performed an 
exploration of the 7 dimensional parameter space of the two-locus, two-allele system 
to determine under which conditions recombination is beneficial 
in terms of our two metrics (\ref{metric1twobit}) and (\ref{metric2twobit}).
 In such a high dimensional space, visualization of the resulting graphs 
requires separation into several distinct cases. We set $p_c=0.5$ in all the 
following as $p_c$ just affects the magnitude of the effects of recombination but not whether it is
beneficial or not as this is controlled by the sign of $\Delta$.
\footnote{Save for the non-generic values $p_c=0$ and $p_c=1$, there are no important qualitative changes as a function of the recombination
probability.} 

\label{results}
\subsection{Recombination as a function of fitness landscape}

We first consider graphs for arbitrary fitness landscapes but for a fixed initial population, with
a further subdivision into cases made according to the type of initial population. As we have fixed $b_1=b_2=b$
and set $a=1$ we display the graphs as functions of $b$ and $c$. The valid region, all fitnesses positive 
with the genotype $11$ as optimum, is given by $b>-1$, $c>-2b$ and $c>-b$. The deceptive region is 
given by $b<0$. For ease of interpretation we also show lines associated with the multiplicative limit 
$b^2=c$ (yellow) and the additive limit $c=0$ (green). Note that both the additive and multiplicative limits 
require $b>0$. The ``needle-in-a-haystack'' landscape is given by $b=0$, $c>0$ and lies on the border
that separates non-deceptive and deceptive landscapes. The point $b=0$, $c=0$ corresponds to a flat
fitness landscape where there is no selection pressure.

Two kinds of graphs are provided, one that displays the value of the SWLD coefficient in different generations, and another 
that displays $\Delta_{ \bar{f}}$ (Equation (\ref{DELTAf})), defined as the change in average fitness between generation 
$t$ and generation $t+1$ in a population evolving with both selection and recombination minus the change in 
average fitness of the same population but evolving with selection only. 
In the graphs we show four representative time slices - $t=1$, 2, 6, and 10 generations
after the initial one. The plane $\Delta_{11}=0$ that separates the recombination 
advantageous/disadvantageous regimes is displayed (turquoise in the online version). For a given generation, those
values of $b$ and $c$ where $\Delta_{11}<1$ are shaded in red (below the $\Delta_{11}=0$ plane),
while those where $\Delta_{11}>1$ correspond to a darker shading (above the $\Delta_{11}=0$ plane). 

\subsubsection{Initial Population $P_{00}\approx 1$ }
\label{initpop00aprox1}

In this first case we consider the dynamics when the initial population is dominated by the non-optimal 
wild type $00$, with $P_{00}(0)=0.8999$, $P_{01}(0)=0.05$, $P_{10}(0)=0.05$, $P_{11}(0)=0.0001$.  
So, we are here interested in the effects of recombination on the dynamics of favourable mutations as 
a function of the fitness landscape and in the background of an initial population dominated by a 
non-optimal wild type. We fix $a=1$ and study the variation in $\Delta$ as a function of $b$ and $c$, 
remembering the restrictions $2b+c > 0$ and $b+c > 0$.  The most notable feature of
\ref{DeltaPZeroOne.eps} is that negative values of $\Delta$ are most associated with additive or 
negatively epistatic landscapes. Note that earlier in the evolution, $t=1$, the benefits of recombination 
are clear to see, even for quite positively epistatic interactions with only deceptive landscapes showing 
a disadvantage. This, however, is due to 
this region being still in the {\it search} regime, as the initial frequency of optimal genotypes was zero.
Gradually, the population moves away from the {\it search} regime and enters the {\it modular}
regime, where we see that it is only for landscapes that are either weakly positively epistatic, additive or
negatively epistatic that recombination is beneficial. Note that the relative benefit of recombination is not
fixed but evolves, thus showing the dependence on the relative frequencies of the different genotypes. 
In terms of BBs, $\Delta$ becomes positive when $P_{11}>P_{1*}P_{*1}$ so, as the frequency
of the optimal type increases, eventually recombination becomes unfavourable relative to selection only,
with the point at which it becomes unfavourable, $P_{11}=P_{1*}P_{*1}$, being dependent on the fitness landscape, as well as
the initial population.
\ImagehereFull{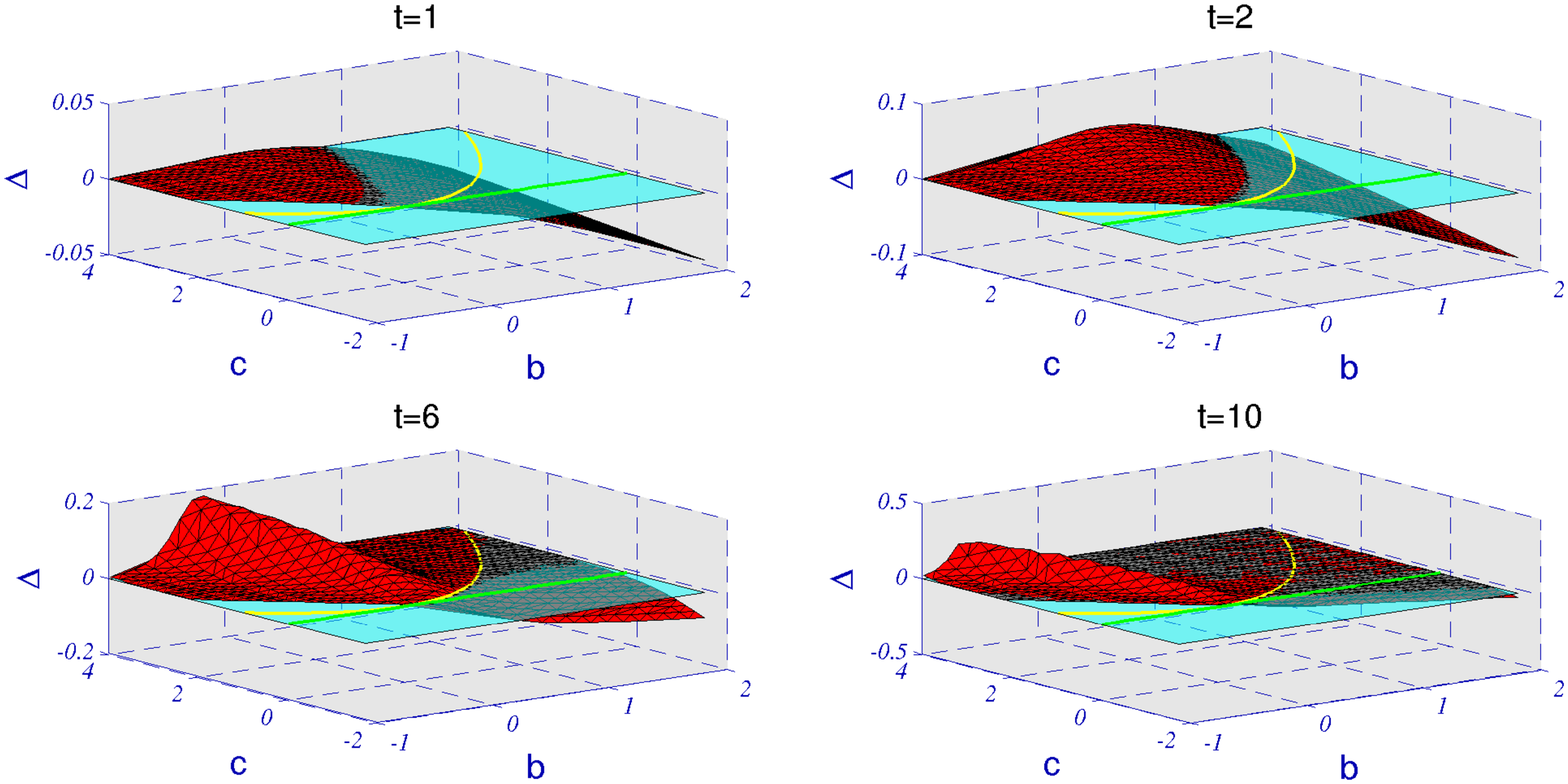}{15.5cm}{ Value of $\Delta$ at different generations for two-locus two-allele system 
as a function of fitness landscape, characterized by $b$ and $c$. The initial population
is $P_{00}(0)=0.8999$, $P_{01}(0)=0.05$, $P_{10}(0)=0.05$, $P_{11}(0)=0.0001$.
The $\Delta=0$ plane has been marked to distinguish 
between conditions in which recombination is favorable ($\Delta<0$) or not. The curve on the plane is $c=b^2$, 
the condition for a multiplicative landscape.}{DeltaPZeroOne.eps}
Turning now to the graphs of the change in average fitness of the population; at $t=1$, in the
{\it search} regime, we see that recombination leads to an increase in average population fitness, over and
above that of selection only, for basically all landscapes. This is due to the addition of optimal genotypes
in an initial population dominated by the non-optimal wild type. Gradually, however the effect of 
recombination diminishes as one enters the {\it modular} regime so that for positively epistatic 
landscapes the difference between selection only and recombinative dynamics is minimal. However,
we note that there is still a strong pronounced effect for either weakly positively epistatic, additive or
weakly negatively epistatic landscapes. 

\ImagehereFull{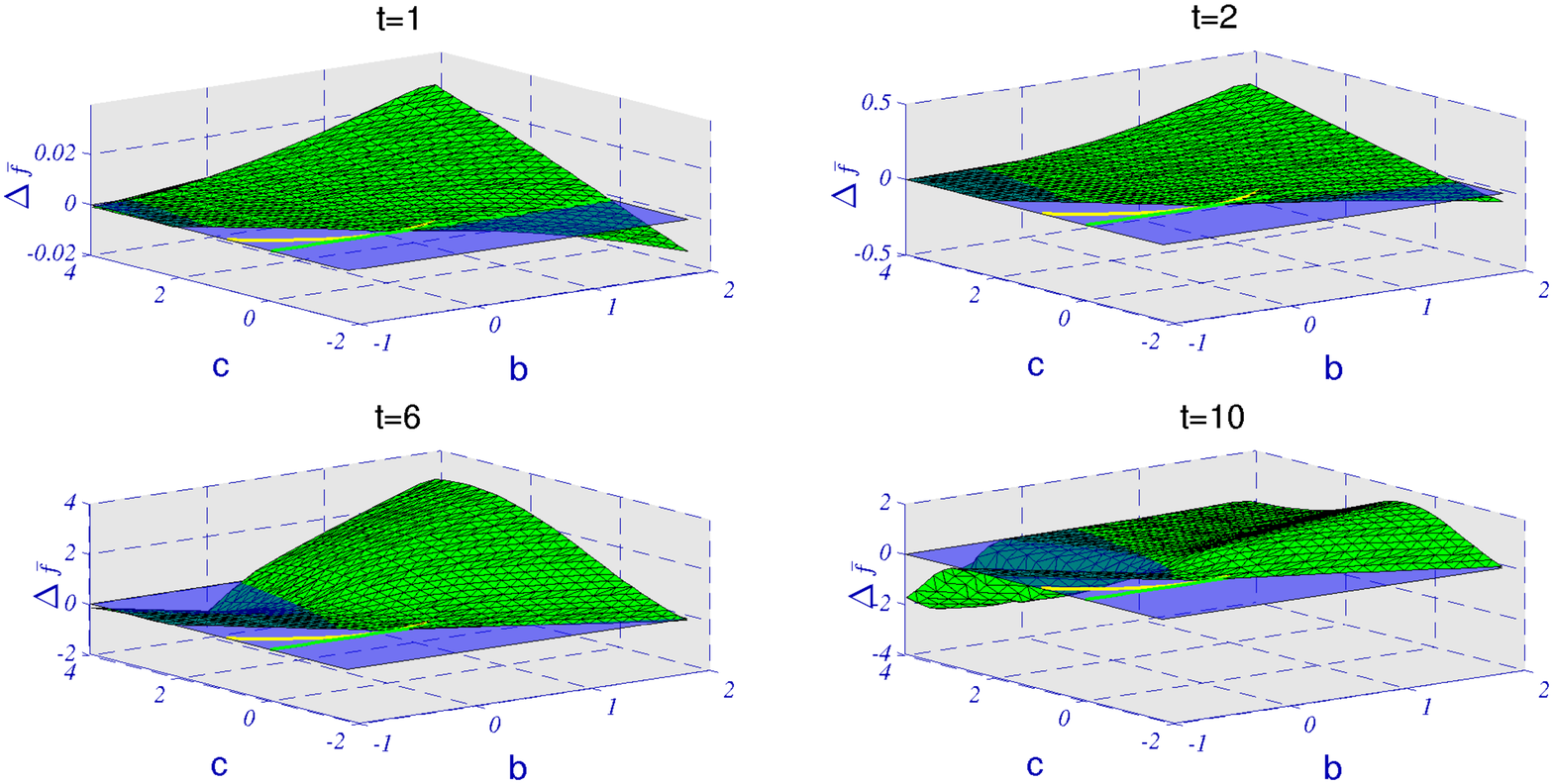}{15.5cm}{ Value of $\Delta_{\bar f}$ at different generations
for the two-locus two-allele system as a function of fitness landscape, characterized by $b$ and $c$. 
The initial population is $P_{00}(0)=0.8999$, $P_{01}(0)=0.05$, $P_{10}(0)=0.05$, $P_{11}(0)=0.0001$.
The $\Delta_{\bar f}=0$ plane has been marked to distinguish between conditions in which 
recombination is favorable ($\Delta_{\bar f}>0$) or not.}{FitnessPZeroOne.eps}

So, how do we interpret these results in terms of BBs? Both in the {\it search} and {\it modular} regimes
the advantage of recombination is associated with the fact that BBs of the optimal genotype, $1*$ and 
$*1$, are recombined to form the type $11$. As the graphs show, this recombination of BBs is, 
in fact, a more efficient process in generating optimal types and increasing overall population fitness 
than selection alone for weakly epistatic landscapes. In fact, the benefit in the {\it search} regime is actually 
relatively independent of the degree of epistasis of the landscape. Later on though, in the 
{\it modular} regime, the generation of optimal genotypes by recombining optimal BBs competes 
against the generation that evolved through pure selection effects. For positively epistatic landscapes, once there
are enough optimal types selection can produce new ones as or more efficiently than recombination.
For {\it modular} landscapes however, recombination retains its advantage. Indeed, this is, in fact, 
what characterizes the {\it modular} regime, i.e., that weakly epistatic BBs or modules are juxtaposed 
by recombination into even fitter genotypes leading to a faster evolution and a faster increase in 
average population fitness.  The fact that the recombination is even more beneficial in the presence of
additive negative epistasis is due to the fact that the destruction of the optimal type produces two
single mutants that have fitness very similar to that of the optimal type. This is the advantage of 
genetic redundancy.

\subsubsection{Initial Population $P_{11}\approx1$ }
\label{initpop11}

We now turn to the case where the initial population is dominated by the optimal genotype as the 
wild type with the presence of genotypes with a single deleterious mutation and a small proportion
of deleterious double mutant genotypes. Specifically, $P_{11}(0)=0.8999$, $P_{10}(0)=0.05$, 
$P_{01}(0)=0.05$ and $P_{00}(0)=0.0001$. The question now is: What is the dynamics of the 
deleterious mutations in the population as a function of the landscape parameters?
Once again, we fix $a=1$ and study the variation in $\Delta$ as a function of $b$ and $c$, 
\ImagehereFull{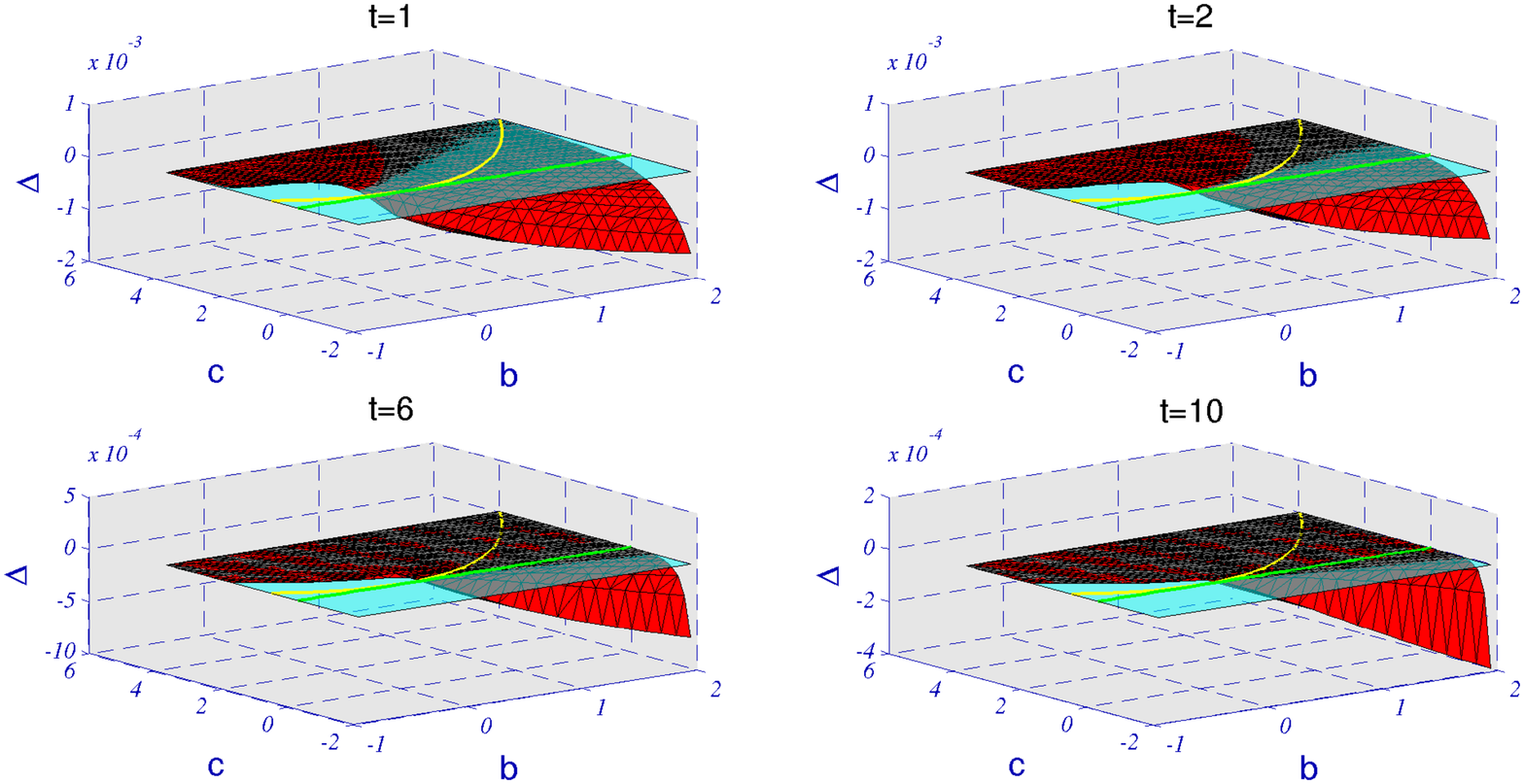}{15.5cm}{Value of $\Delta$ at different generations for 
two-locus two-allele system as a function of fitness landscape, characterized by $b$ and $c$. 
The initial population is $P_{11}(0)=0.8999$, $P_{10}(0)=0.05$, $P_{01}(0)=0.05$ and 
$P_{00}(0)=0.0001$. The $\Delta=0$ plane has been marked to distinguish 
between conditions in which recombination is favorable ($\Delta<0$) or not. The curve 
on the plane is $c=b^2$, the condition for a multiplicative landscape.}{DeltaPOneOne.eps}

In Figure \ref{DeltaPOneOne.eps} the first thing to notice is that, in distinction to the case where the
initial population is dominated by the non-optimal genotype, here there is no dinstinct behavior associated
with the {\it search} regime, as the optimal genotype is already dominant in the population. Thus,
for positively epistatic landscapes the difference due to recombination is small. However, for
additive or negatively epistatic landscapes we see that recombination is advantageous, with 
the advantage being more significant in the presence of negative epistasis. This is due to the 
fact that in such landscapes the elimination of the suboptimal double mutant $00$ is more efficient. 
\ImagehereFull{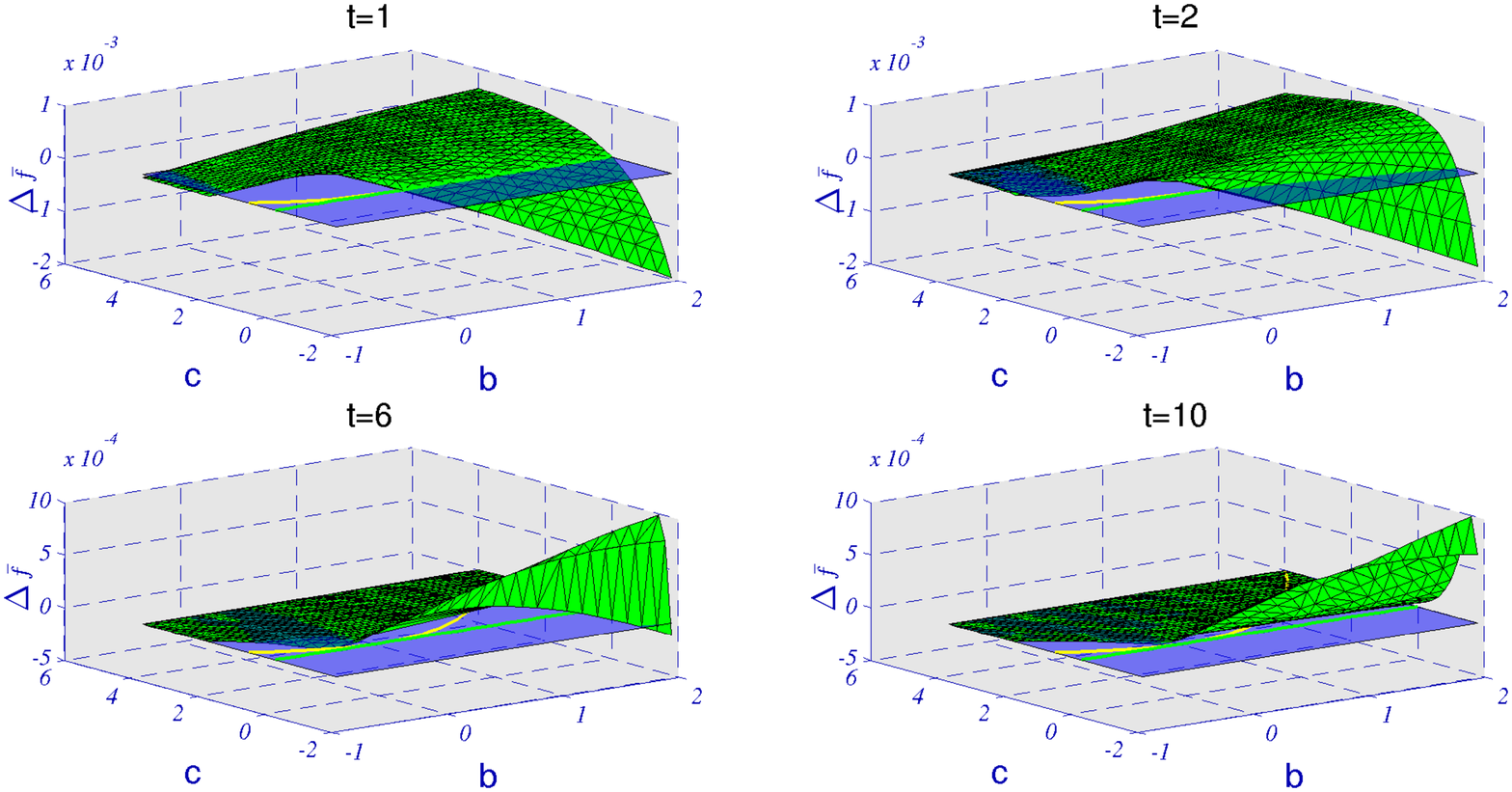}{15.5cm}{Value of $\Delta_{\bar f}$ at different generations
for the two-locus two-allele system as a function of fitness landscape, characterized by $a$ and $c$. 
The initial population is $P_{11}(0)=0.8999$, $P_{10}(0)=0.05$, $P_{01}(0)=0.05$ and 
$P_{00}(0)=0.0001$. The $\Delta_{\bar f}=0$ plane has been marked to distinguish between 
conditions in which recombination is favorable ($\Delta_{\bar f}>0$) or not.}{FitnessPOneOne.eps}

Considering now the average population fitness, we see clearly in Figure \ref{FitnessPOneOne.eps} 
how the advantage of recombination manifests itself in the {\it modular} regime where epistasis is
weak. Interestingly, we see how negatively epistatic landscapes are, in the early part of the evolution,
associated with $\Delta_{\bar f}<0$. This is due to the fact that for negative epistasis the overall 
contribution to the population fitness of a deleterious double mutant and an optimal genotype is less
double mutant, selection can eliminate the mutations thereby purifying the population more efficiently 
than selection alone. The more modular the landscape the more efficient this process becomes. 

\subsubsection{Initial Population $P_{11}\approx0$, $P_{00}\approx\frac{1}{2}$,
$P_{01}\approx P_{10}\approx \frac{1}{4}$ }
\label{initpop00half}

We now consider a scenario similar to that of sub-section \ref{initpop00aprox1}, where the 
initial proportion of optimal genotypes is zero; but now, however, the frequency 
of the BBs, $1*$ and $*1$, represented by the beneficial mutants $01$ and $10$, relative to the 
less fit wild type $00$ is much higher. Concretely, the initial population is: 
$P_{11}(0)=0.0001$, $P_{10}(0)=0.25$, $P_{01}(0)=0.25$ and $P_{00}(0)=0.4999$ so that 
the BBs $1*$ and $*1$ form about a quarter of the population each one. 
\ImagehereFull{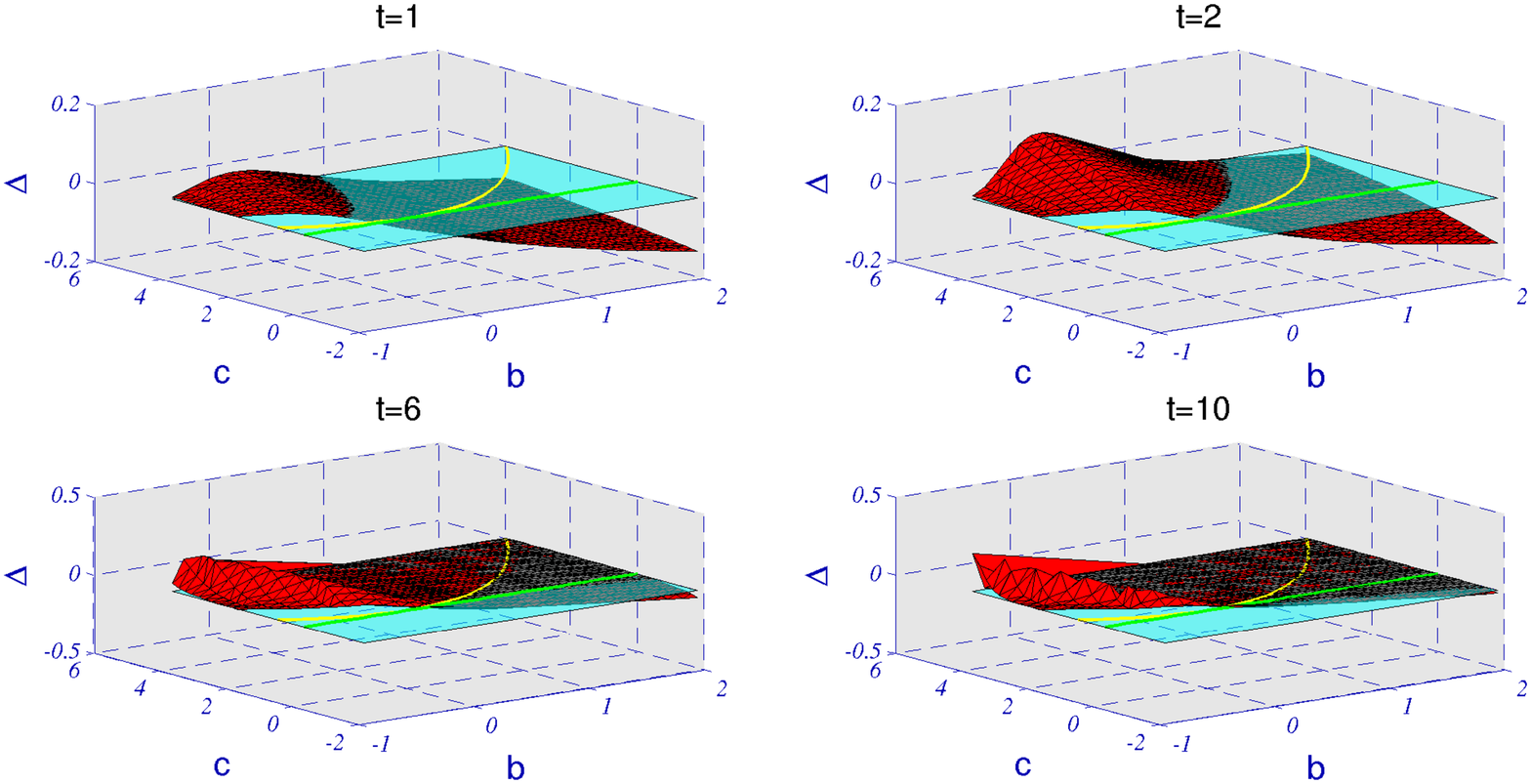}{15.5cm}{Value of $\Delta$ at different generations for two-locus 
two-allele system as a function of fitness landscape, characterized by $b$ and $c$. The initial population
is $P_{00}(0)=0.4999$, $P_{01}(0)=0.25$, $P_{10}(0)=0.25$, $P_{11}(0)=0.0001$.
The $\Delta=0$ plane has been marked to distinguish between conditions in which recombination 
is favorable ($\Delta<0$) or not. The curve on the plane is $c=b^2$, the condition for a multiplicative 
landscape.}{DeltaPOneZero.eps}

We see in Figure \ref{DeltaPOneZero.eps} that the graphs are qualitatively similar to those of 
Figure \ref{DeltaPZeroOne.eps}. The chief difference now is that recombination is even more disadvantageous
in the {\it search} regime for deceptive landscapes than before and more advantageous for modular 
landscapes - weak or zero positive epistasis or negative epistasis. This is due to the wider availability of the BBs $1*$ and $*1$
thus obstructing/facilitating the construction of the optimal type $11$ according to whether the landscape
is deceptive or modular. As evolution progresses, as before,
we see a passage from the {\it search} regime to the {\it modular regime}, where the relative benefit 
of recombination is restricted to weakly positively epistatic, additive or negatively epistatic landscapes. 

\ImagehereFull{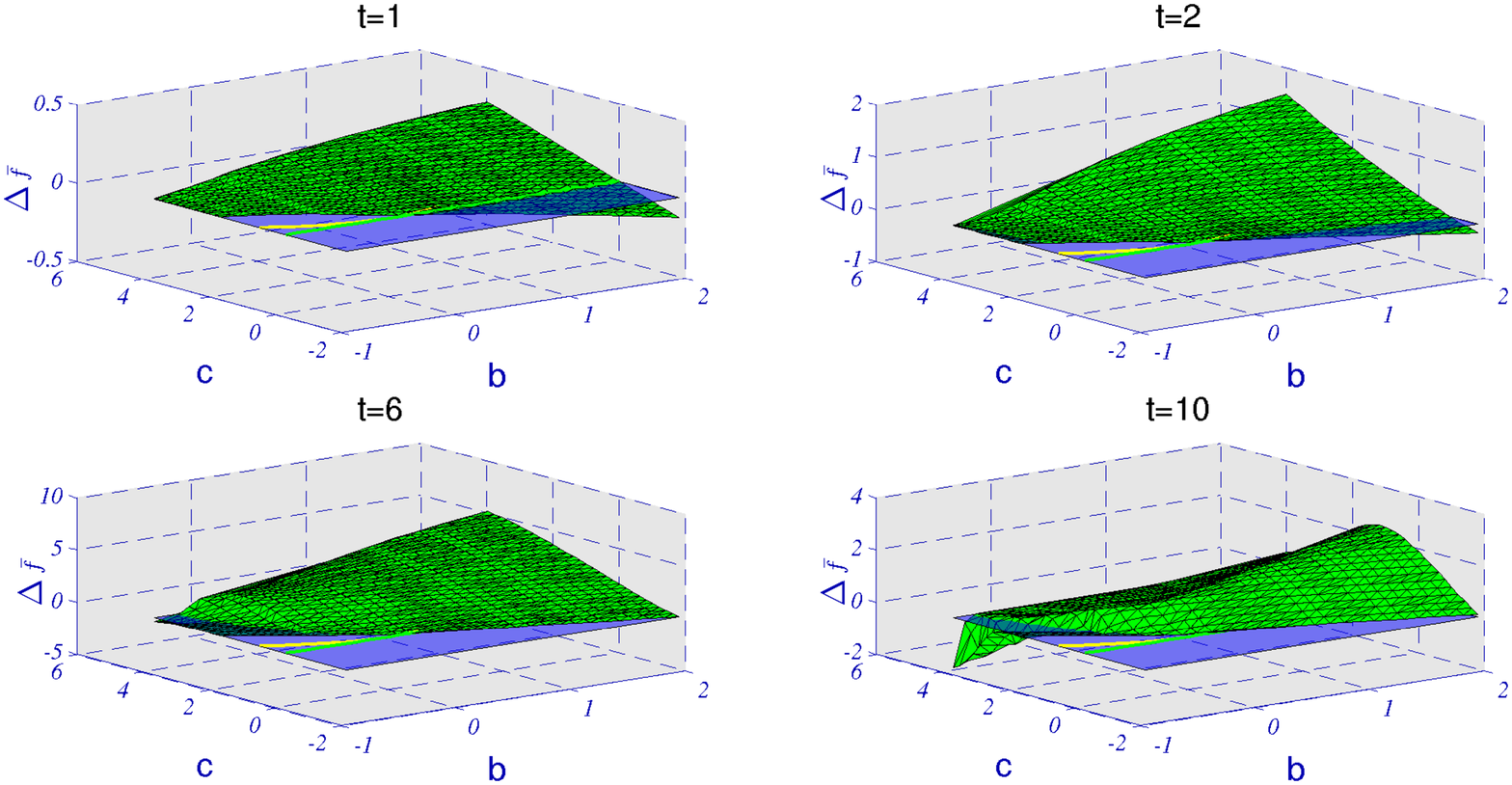}{15.5cm}{ Value of $\Delta_{\bar f}$ at different generations
for the two-locus two-allele system as a function of fitness landscape, characterized by $b$ and $c$. 
The initial population is $P_{11}(0)=0.0001$, $P_{10}(0)=0.25$, $P_{01}(0)=0.25$ and 
$P_{00}(0)=0.4999$. The $\Delta_{\bar f}=0$ plane has been marked to distinguish between 
conditions in which recombination is favorable ($\Delta_{\bar f}>0$) or not.}{FitnessPOneZero.eps}

Similarly, in Figure \ref{FitnessPOneZero.eps} we see a similarity with the corresponding graphs of 
Figure \ref{FitnessPZeroOne.eps} the average population fitness showing a strong increase, relative
to the selection only case, due to the efficient formation of the optimal type, which in its turn is due to 
the large number of BBs in the population.  Even for strongly epistatic landscapes there is a strong 
benefit to recombination in this regime. At later times, in the {\it modular} regime, we see that the 
advantage of recombination is again associated with additive, weakly positively epistatic or negatively epistatic 
landscapes, i.e., modular landscapes. 

So, we see that the principle effect of increasing the BB frequency in the initial population is to 
accelerate the rate of evolution so that the frequency of the optimal genotype and the average 
population fitness increase more rapidly.

\subsubsection{Initial Population $P_{11}\approx0$, $P_{00}\approx0$ }
\label{initpop11000}

We now look at an even more extreme case, where the initial population is completely dominated 
by the single mutants $01$ and $10$ with the initial population being $P_{11}(0)=0.0001$, 
$P_{10}(0)=0.4999$, $P_{01}(0)=0.4999$ and $P_{00}(0)=0.0001$. Qualitatively the results are
as in sub-sections \ref{initpop00half} and \ref{initpop00aprox1}; the strong presence of the 
BBss $1*$ and $*1$ leading to a very efficient production of the optimal genotype $11$. 
This is, in fact, another good illustration of Muller's ratchet. Although recombination leads to 
the generation of optimal genotypes it also leads to the production of the sub-optimal double
mutants $00$. The latter, however, as the graphs clearly show, are flushed out by selection.
In fact, as Figure \ref{DeltaPOneZeroZero.eps} shows, they are produced and then flushed out most efficiently 
in the presence of recombination for modular landscapes when compared to selection only.

\ImagehereFull{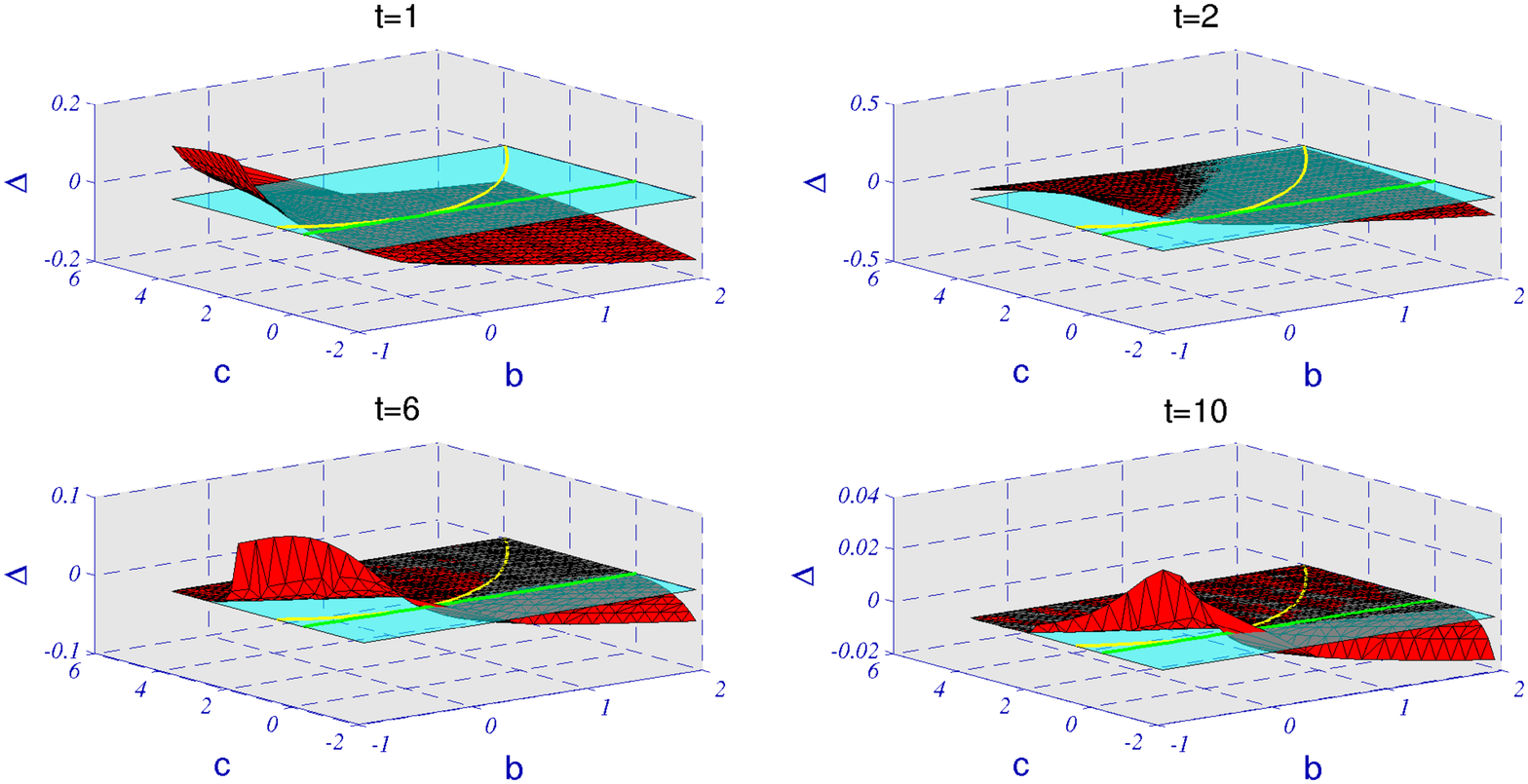}{15.5cm}{Value of $\Delta$ at different generations for two-locus 
two-allele system as a function of fitness landscape, characterized by $b$ and $c$. The initial population
is $P_{11}(0)=0.0001$, $P_{10}(0)=0.4999$, $P_{01}(0)=0.4999$ and $P_{00}(0)=0.0001$.
The $\Delta=0$ plane has been marked to distinguish between conditions in which recombination is 
favorable ($\Delta<0$) or not. The curve on the plane is $c=b^2$, the condition for a multiplicative landscape.}{DeltaPOneZeroZero.eps}
\ImagehereFull{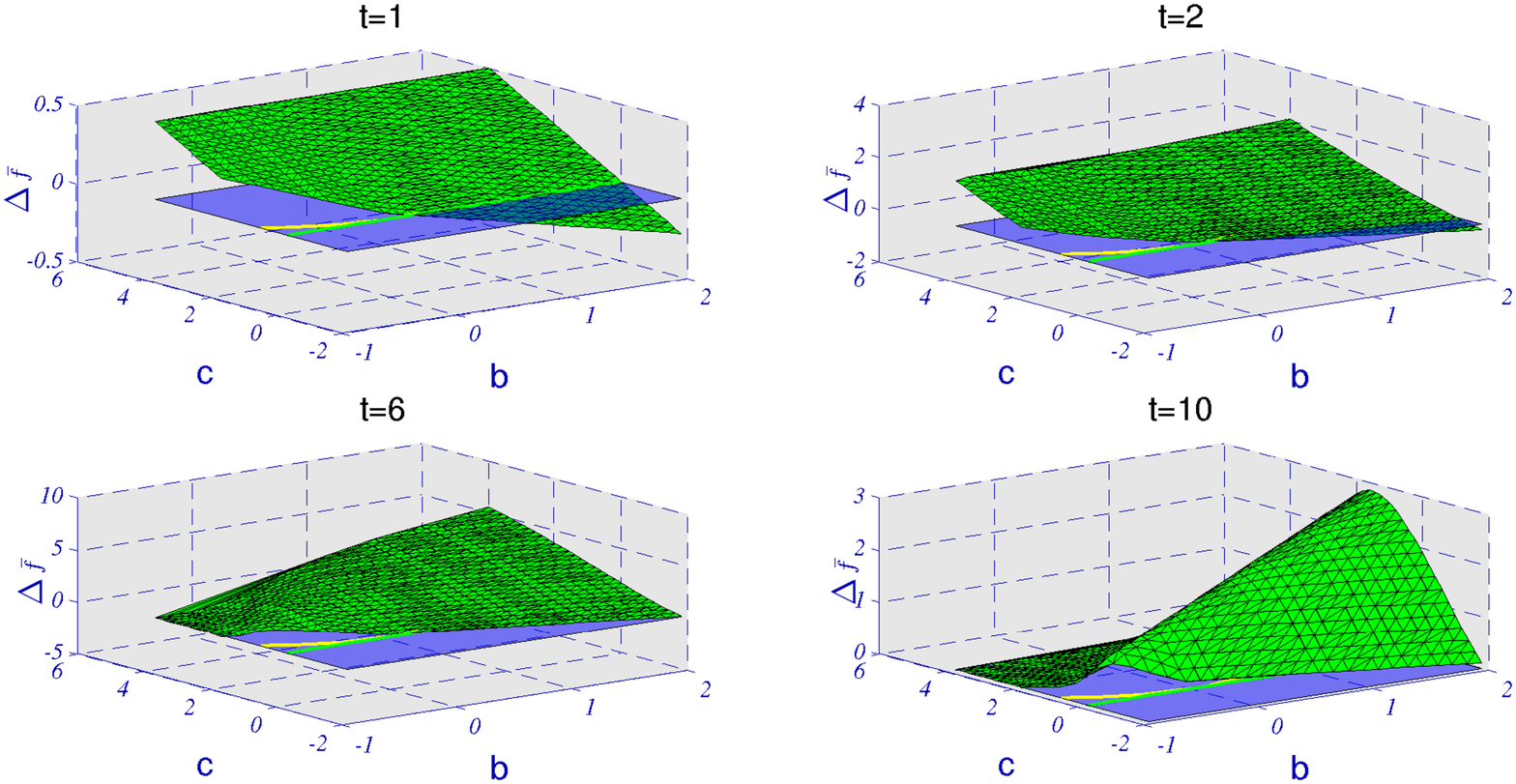}{15.5cm}{ Value of $\Delta_{\bar f}$ at different generations
for the two-locus two-allele system as a function of fitness landscape, characterized by $b$ and $c$. 
The initial population is $P_{11}(0)=0.0001$, $P_{10}(0)=0.4999$, $P_{01}(0)=0.4999$ and 
$P_{00}(0)=0.0001$. The $\Delta_{\bar f}=0$ plane has been marked to distinguish between 
conditions in which recombination is favorable ($\Delta_{\bar f}>0$) or not.}{FitnessPOneZeroZero.eps}
\subsubsection{Initial Homogeneous Population $P_{ij}=0.25$}
\label{initpophomogeneous}

The final initial population type we will consider is that of a uniform initial population where all
genotypes have the same initial frequency, $0.25$. Here we see behaviour that is qualitatively similar 
to that found for other populations. The chief difference here is that given the ample presence 
of the optimal genotype in the initial population there is no {\it search} regime and so the dynamics 
begins and remains in the {\it modular} regime. With no population bias we can see the role played
by the multiplicative limit with at $t=1$ $\Delta$ being positive for landscapes with positive multiplicative
epistasis and, particularly, deceptive landscapes. It is negative for weakly postively epistatic, additive and
negatively epistatic landscapes. As evolution progresses we can see that the relative advantage diminshes such that 
at $t=10$ the advantage of recombination is only noticeable for larger negative epistasis.
\ImagehereFull{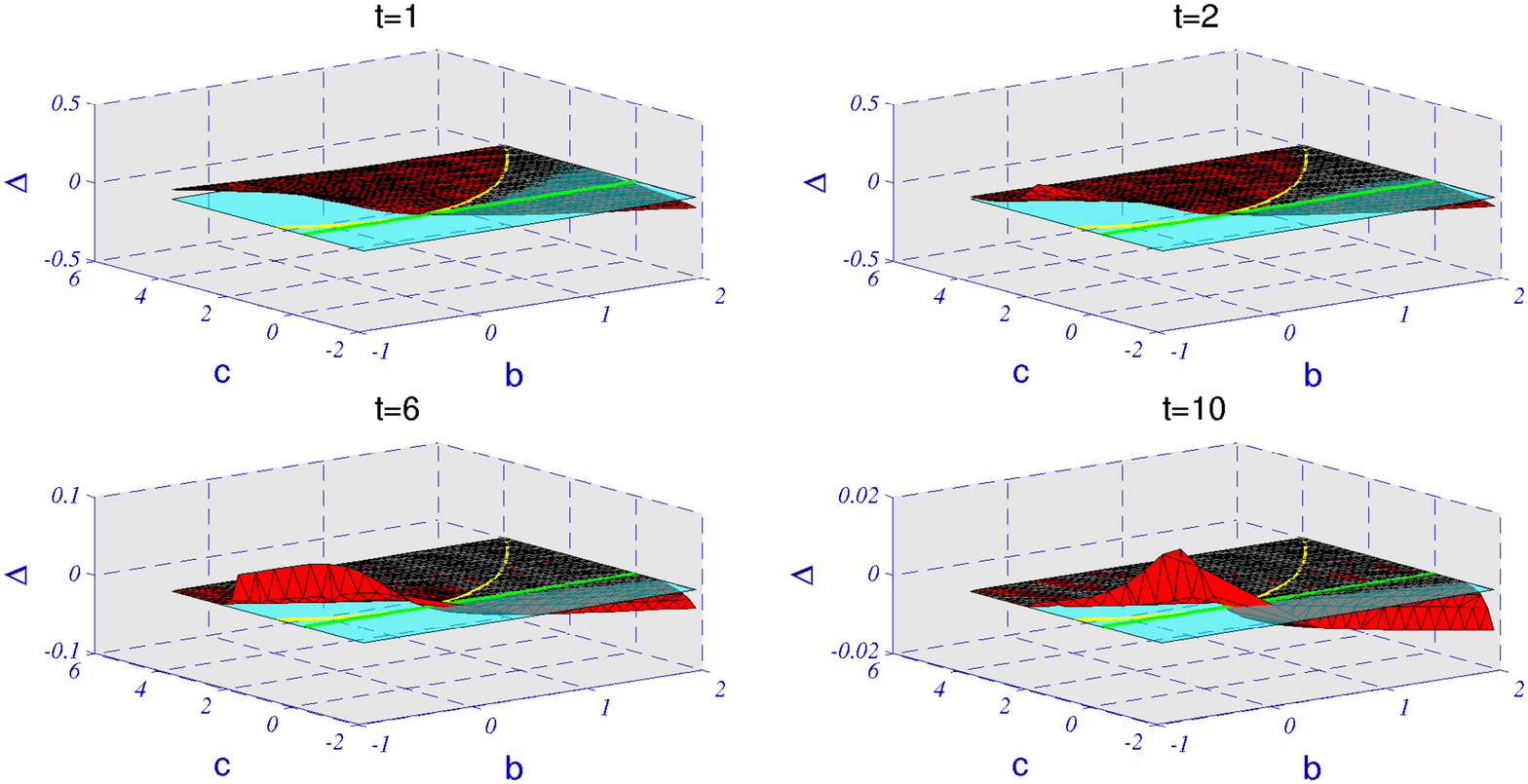}{15.5cm}{Value of $\Delta$ at different generations for 
two-locus two-allele system as a function of fitness landscape, characterized by $b$ and $c$. The 
initial population is $P_{ij}(0)=0.25$. The $\Delta=0$ plane has been marked to distinguish 
between conditions in which recombination is favorable ($\Delta<0$) or not. The curve on the plane is 
$c=b^2$, the condition for a multiplicative landscape.}{DeltaHomogeneous.eps}
\ImagehereFull{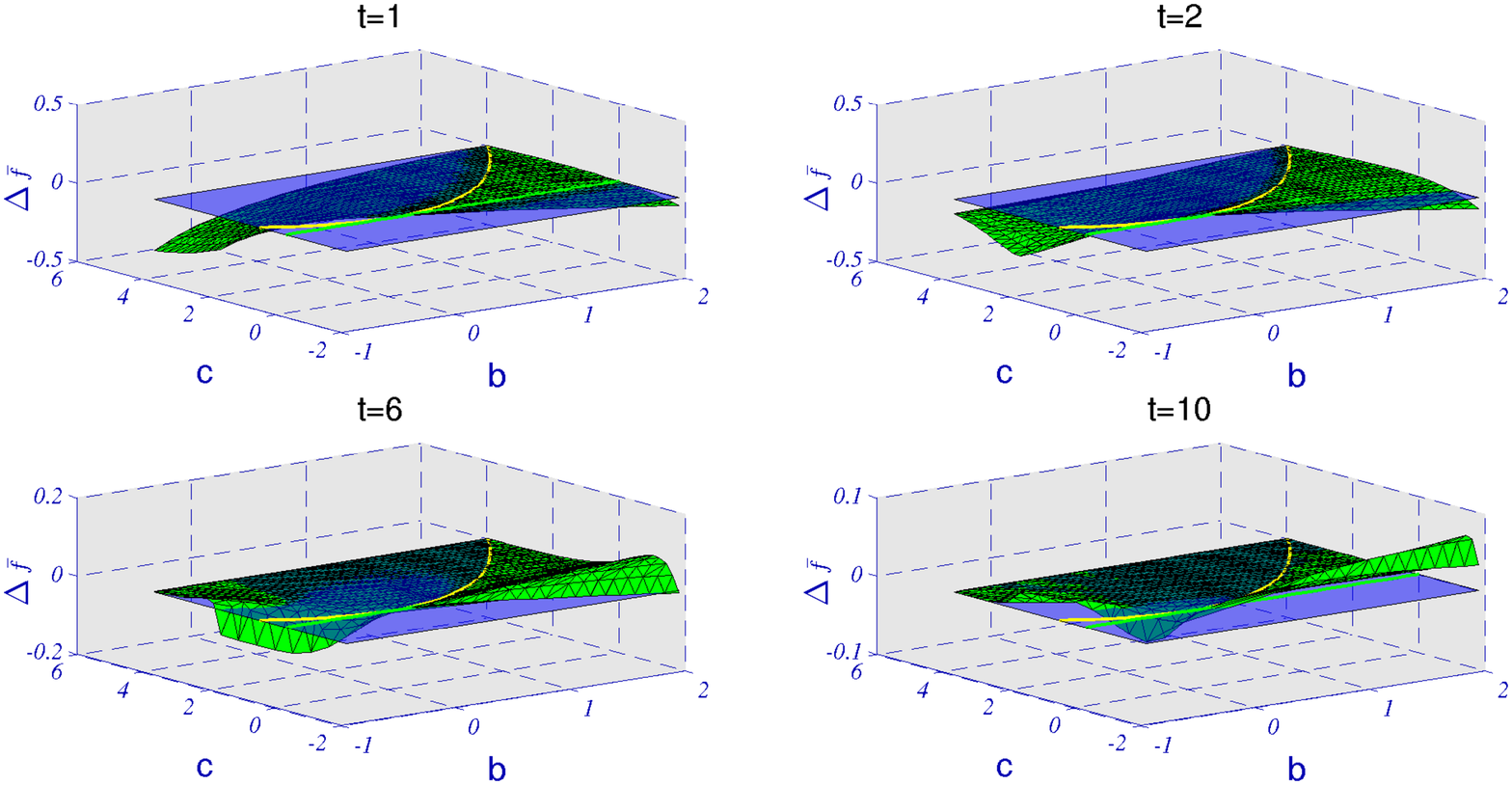}{15.5cm}{ Value of $\Delta_{\bar f}$ at different generations
for the two-locus two-allele system as a function of fitness landscape, characterized by $b$ and $c$. 
The initial population is $P_{ij}(0)=0.25$. The $\Delta_{\bar f}=0$ plane has been marked to 
distinguish between conditions in which recombination is favorable ($\Delta_{\bar f}>0$) or not.}{FitnessHomogeneous.eps}
In terms of average population fitness in Figure \ref{FitnessHomogeneous.eps} we see an analogous
story: at $t=1$ average population fitness is increased only for landscapes with negative multiplicative 
epistasis, up to the additive limit, but is, in fact, negative for negative additive epistasis. However, as evolution
progresses, once again, we see the dominant role played by modular landscapes - i.e., weakly positively epistatic,
additive and negatively epistatic landscapes.

\subsection{Recombination as a function of population}

Having explored the effect of recombination on the space of fitness landscapes, by varying 
continuously the landscape parameters $b$ and $c$ for a variety of distinct initial populations, 
we now consider the complementary viewpoint of considering how the effect of recombination
changes by varying continuously the initial population for a variety of fixed fitness landscapes.
Due to the conservation of probability, the population vector is characterized by only three frequencies.
For simplicity of visualization we will consider intitial populations such that $P_{01}(0)=P_{10}(0)$
and consider the population dynamics as a function of $P_{11}(0)$ and $P_{01}(0)$. 

A general observation on almost all the graphs in this section is that since there is generic convergence to the
optimal genotype  $P_{11}=1$ for non-deceptive landscapes so clearly all the surfaces have 
$\Delta =0$ in the $P_{11}=1$ corner. 

\subsubsection{Additive landscape $a=1, b_1=b_2=1, c=0$.}

The first landscape we will consider is an additive landscape ($c=0$). 
For this landscape (Figure \ref{CODelta.eps}) the tendency is clear, that the more BBs and the fewer 
optimal types there are, the more recombination helps. This is again a manifestion of the {\it search} regime. 
In this landscape, as can be seen at $t=1$, recombination in terms of $\Delta$ is only 
unfavorable when the proportion of optimal types is appropriately larger than the frequencies 
of the BBs, as then selection can act more efficiently to increase the frequency of the optimal 
type than can recombination of the single mutants. However, we see that this effect is temporary. 
By $t=6$ basically any initial population is associated with $\Delta < 0$. We can see 
that the SWLD increases in time, approaching zero asymptotically, this regime being 
associated with the approach to a population completely dominated by the optimal genotype.  
This dynamics, in fact, shows an important universality associated with recombination, 
that demonstrates the role of Muller's ratchet: that the action of recombination is
to drive the system to particular frequencies for the optimal type and its BBs that correspond to 
quite special initial conditions at $t=0$. To understand this, note that at $t=6$ 
and $t=10$ the proportion of optimal types is high. If we imagine the value of $P_{11}(t=6)$,
for example, that is a consequence of evolution in the presence of recombination, then we can
map those values such as to imagine them as initial conditions, say at $t=1$, for further evolution.  
However, we can observe at $t=1$ that values of $P_{11}$ close to 1 correspond to positive 
values of $\Delta$ except in a very narrow wedge where the values of $P_{01}$ are as high as possible.
This wedge is associated precisely with a lower relative frequency of the suboptimal $00$ genotype.
The conclusion is that recombination is removing the suboptimal $00$ genotype more efficiently 
than selection only. 
\ImagehereFull{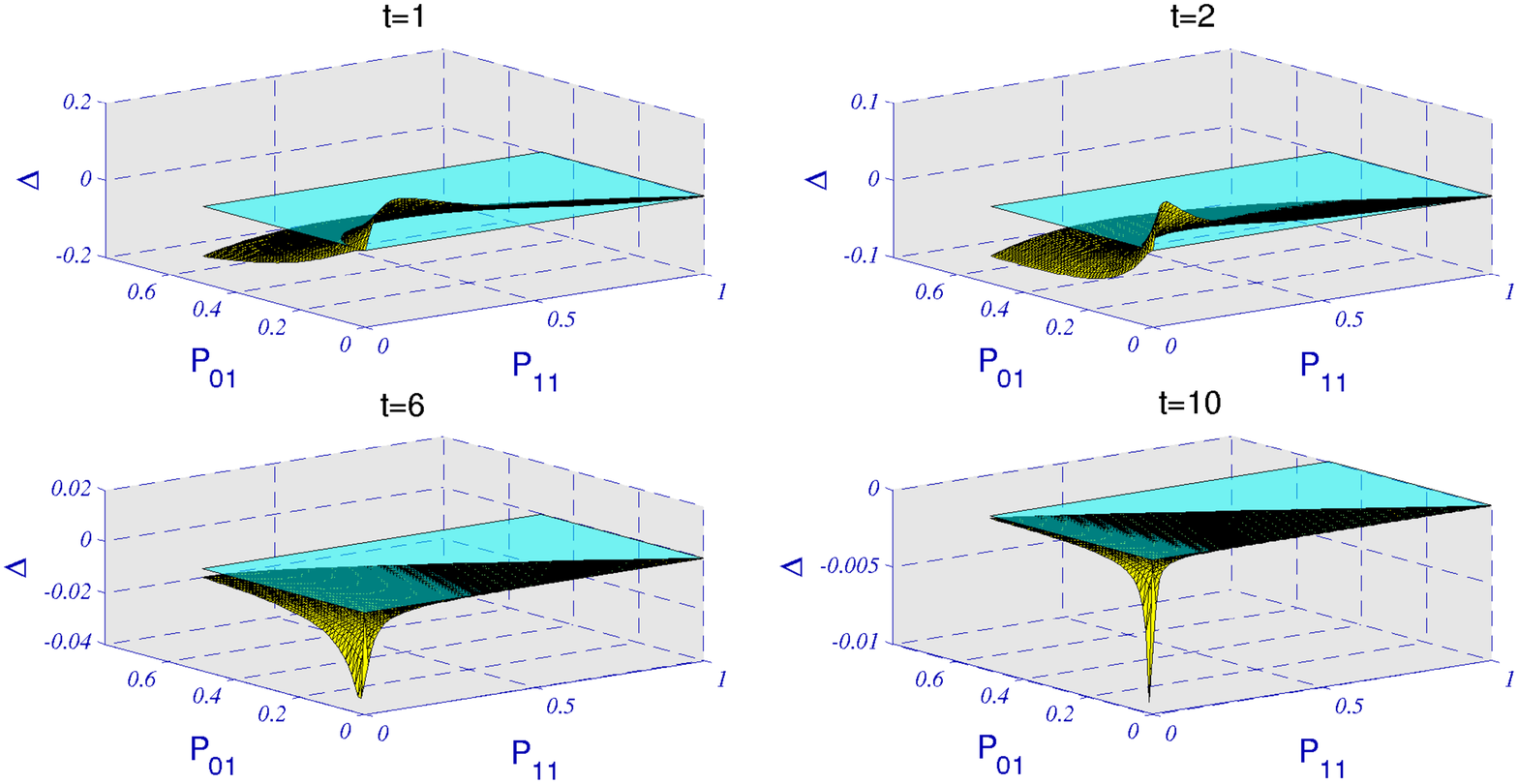}{13.5cm}{Value of $\Delta$ at different time steps for a two-locus 
two-allele system with an additive fitness landscape $a=1, b_1=b_2=1, c=0$) for different values of 
the initial population given by $P_{11}$ and $P_{10}(=P_{01})$. The $\Delta=0$ plane has been marked to 
distinguish between conditions in which recombination is favorable ($\Delta<0$) or not.}{CODelta.eps}

Finally, the presence of a trough associated with quite negative values of $\Delta$
for $t=6$ and $t=10$ is 
a consequence of th fact that the search regime is more extensive when the frequency of both 
optimal genotype and BBs is low.

\subsubsection{Neutral landscape: $b_1=b_2=c=0$, $a\neq0$ ($A=1$)}
\label{Flat}

For a neutral landscape, where the effects of selection are null, as with the additive landscape, 
the ``the more BBs the better recombination is'' rule is valid, but we see a different 
behavior as a function of initial population. For neutral evolution, the SWLD, $\Delta$, and the 
standard linkage disequilibrium coefficient, $D$, are the same. So, Figure \ref{FlatDelta.eps} shows the 
approach to the Geiringer or Robbins manifold, defined by $D=0$. The approach to this manifold 
is from the negative or positive side depending on whether the initial population is dominated
by the BBs $01$ and $10$, or by the optimal genotype $11$. The Geiringer limit has been amply
studied in the literature \cite{geiringer44}. Thus, recombination is beneficial
when there is an ample supply of BBs and few optimal types, and deleterious when there are 
no BBs. The minimal value of $\Delta$ is for $P_{01}(0)=0.5$ and the maximal for $P_{01}(0)=0$, 
$P_{00}(0)=P_{11}(0)=0.5$.
\ImagehereFull{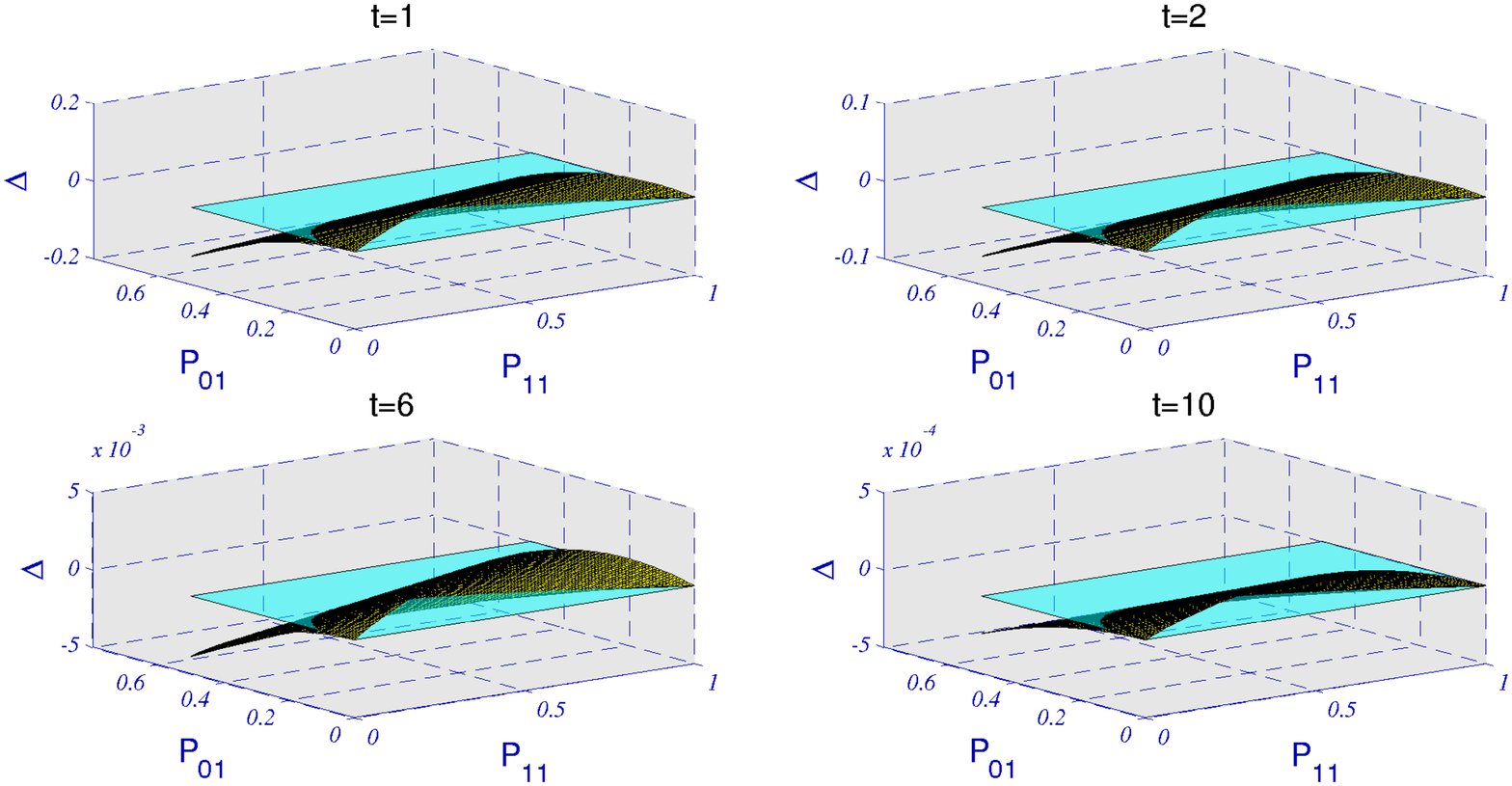}{13.5cm}{Value of $\Delta$ at different time steps for a two-locus 
two-allele system with a neutral ($b_1=b_2=c=0$, $a\neq0$) fitness landscape for different values of 
the initial population given by $P_{11}(0)$ and $P_{10}(0)=P_{01}(0)$. The $\Delta=0$ plane has been 
marked to distinguish between conditions in which recombination is favorable ($\Delta<0$) or not.}{FlatDelta.eps}
%
%
%
%
\subsubsection{Multiplicative landscape $a=1$, $b_1=b_2=2$, $c=4$}

This landscape satisfies the multiplicative constraint that $ac=b^2$. 
Here we see that recombination is favorable in the {\it search} regime where the BB frequency is high
and the frequency of the optimal genotype is low. However, for other than very small $P_{11}$ we can 
see that recombination is somewhat unfavorable when the BB frequency is relatively low but, in the main,
it is generally neutral in its effects. This is consistent with known results for multiplicative landscapes.
In fact, viewing the time evolution, even if one starts in the {\it search} regime we see that very quickly
the system approaches linkage equilibrium. 
\ImagehereCaption{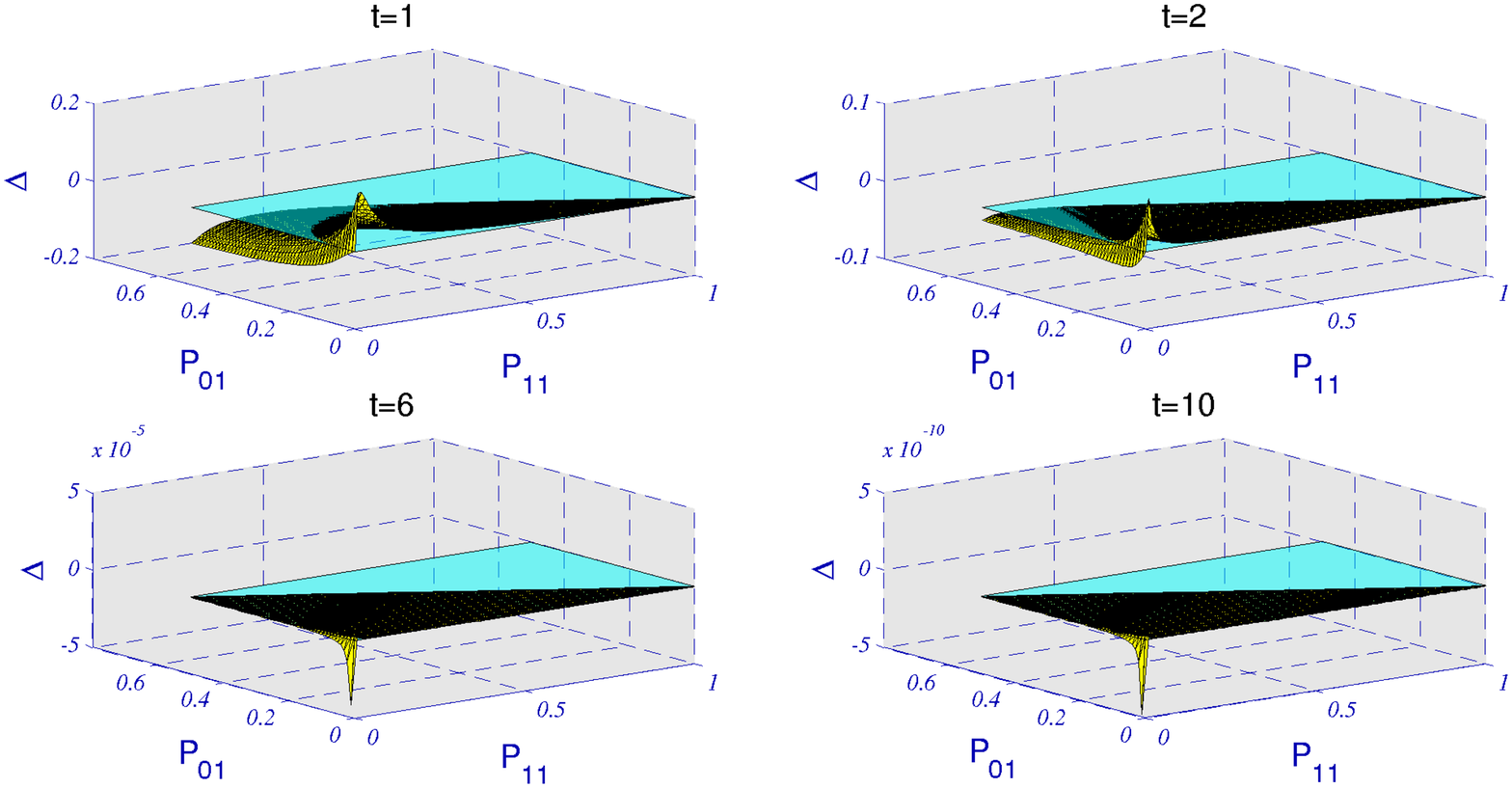}{13.5cm}{Value of $\Delta$ at different time steps for a 
two-locus two-allele system with a multiplicative fitness landscape ($a=1$, $b_1=b_2=2$, $c=4$) 
for different values of the initial population given by $P_{11}$ and 
$P_{10}(=P_{01})$. The $\Delta=0$ plane has been marked to distinguish between conditions in 
which recombination is favorable ($\Delta<0$) or not.}
\subsubsection{Needle-In-A-Haystack, $b_1=b_2=0$, $c\neq0$, $a\neq0$ ($A=\frac{a}{a+c}$) }

We now turn to the case of a landscape with maximally positive epistasis -  NIAH, which, as mentioned, 
has been used extensively in models of molecular evolution and, especially, in considerations of 
selection-mutation balance
and the existence of error thresholds.  Here, it corresponds to a Boolean ``AND" function on the two loci.
As a function of the initial population we can clearly see
that in the {\it search} regime, where there is an ample supply of BBs and only a zero or small 
proportion of the optimal genotype, that recombination is favorable, both in terms of leading to 
a more efficient production of the optimal genotype when compared to selection only ($\Delta < 0$) 
as well as a more fit population ($\Delta_{\bar f}>0$, Figure \ref{NIAHFitness.eps}). On the other hand, away from the {\it search} 
regime it is clear that the effects of recombination are unfavorable. Note that the advantage or 
disadvantage of recombination decreases in time as the system gets closer to linkage equilibrium,
this equilibrium being associated with a population dominated by the optimal genotype.

\ImagehereFull{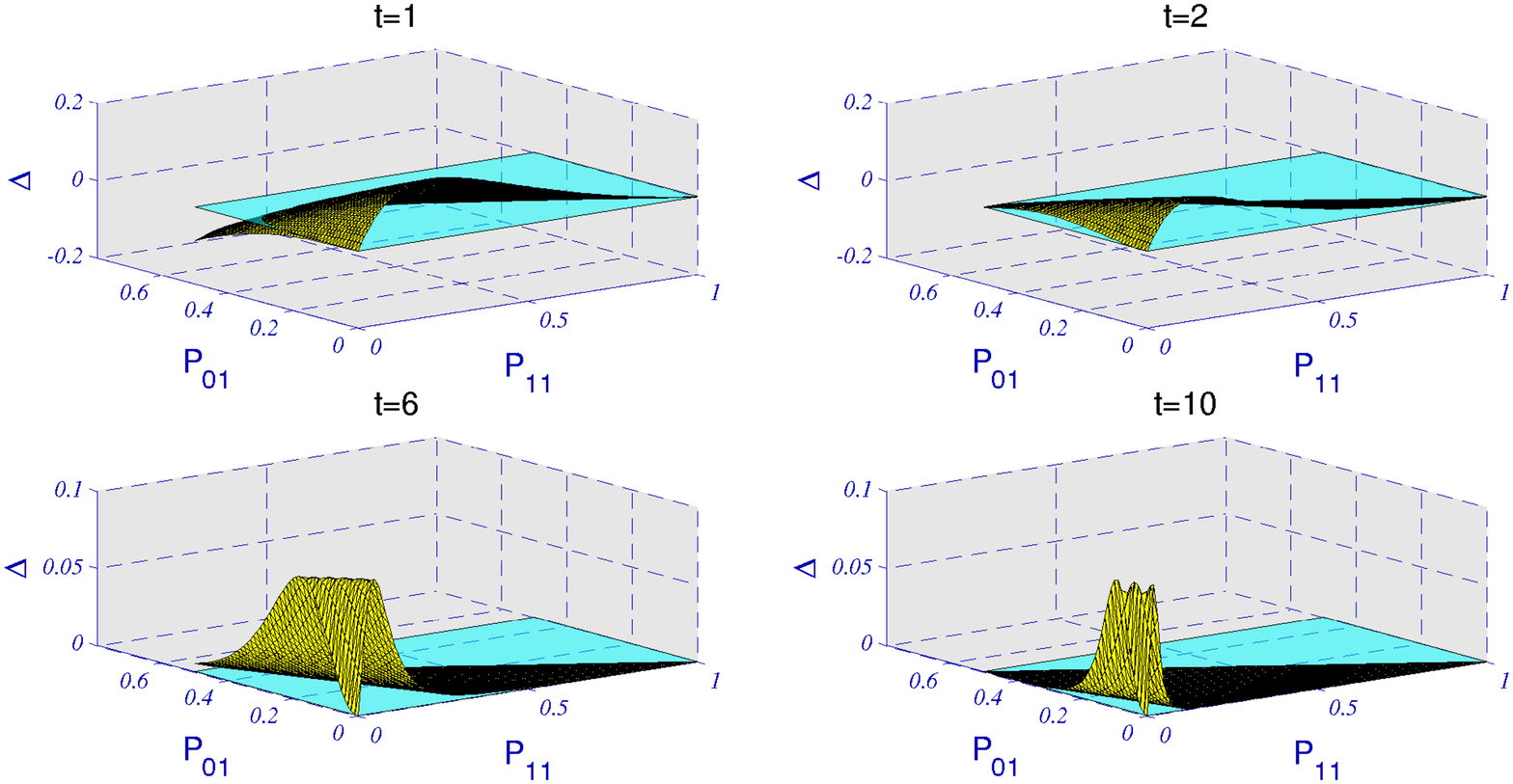}{13.5cm}{Value of $\Delta$ at different generations for a two-locus 
two-allele system with a ``Needle in a haystack'' fitness landscape ($b_1=b_2=0$, $c=0.001$, $a=1$) 
for different values of the initial population given by $P_{11}$ and $P_{10}(=P_{01})$. The 
$\Delta=0$ plane has been marked to distinguish between conditions in which recombination 
is favorable ($\Delta<0$) or not.}{NIAHDelta.eps}

\ImagehereFull{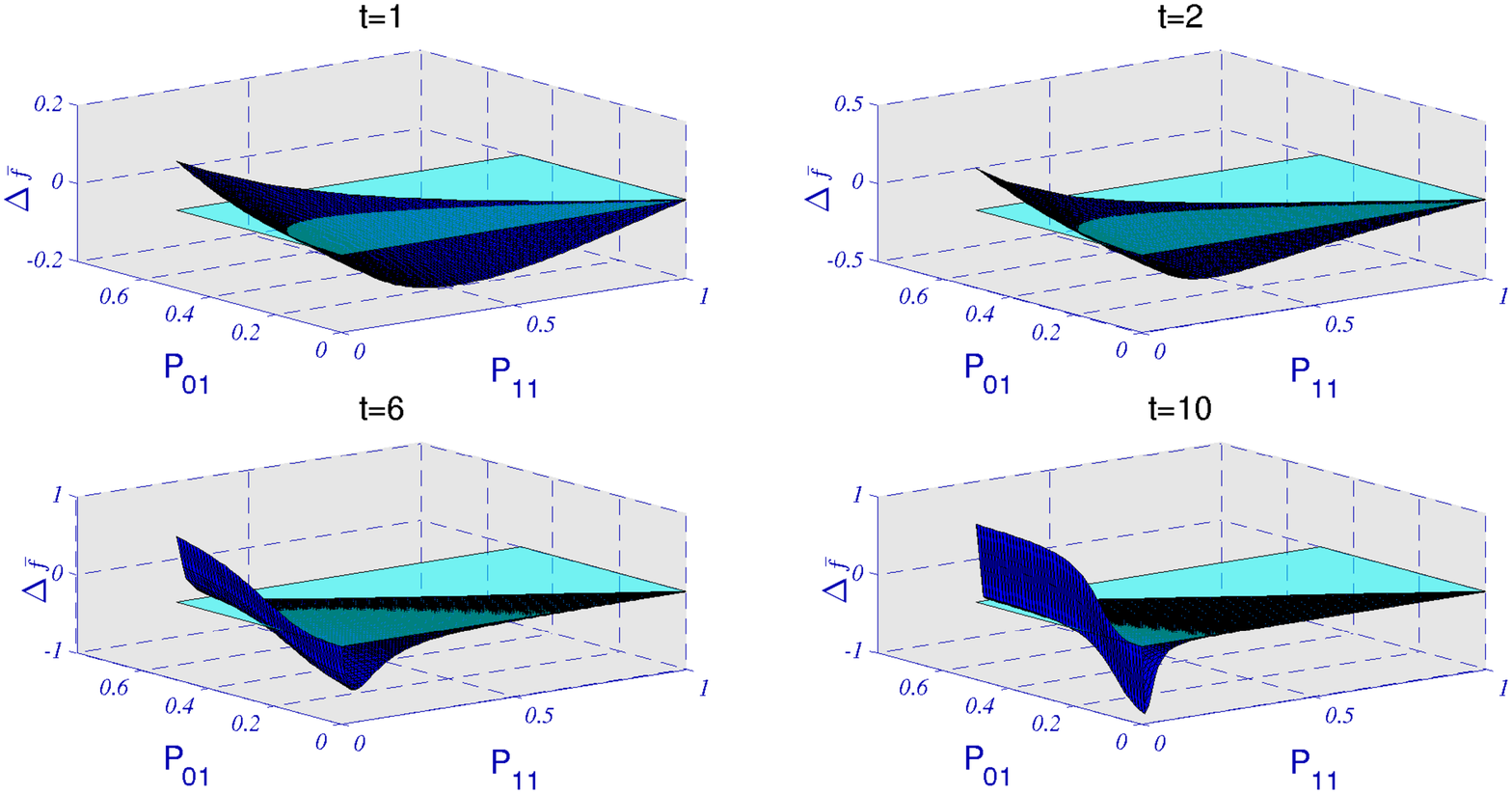}{13.5cm}{Value of $\Delta_{\bar f}$ at different generations for a 
two-locus two-allele system with a ``Needle in a haystack'' fitness landscape ($b_1=b_2=0$, 
$c=0.001$, $a=1$) for different values of the initial population given by $P_{11}$ and $P_{10}(=P_{01})$. 
The $\Delta_{\bar f}=0$ plane has been marked to distinguish between conditions in which 
recombination is favorable ($\Delta_{\bar f}>0$) or not.}{NIAHFitness.eps}

\subsubsection{Landscape with Genetic Redundancy, $a=1$, $b=1$, $c=-1$ }

For a landscape with maximal negative epistasis, corresponding to an "OR" Boolean function on the two
loci we see in Figure \ref{ORDelta.eps} that very rapidly recombination becomes beneficial in terms of $\Delta$
for any initial population.  

\ImagehereFull{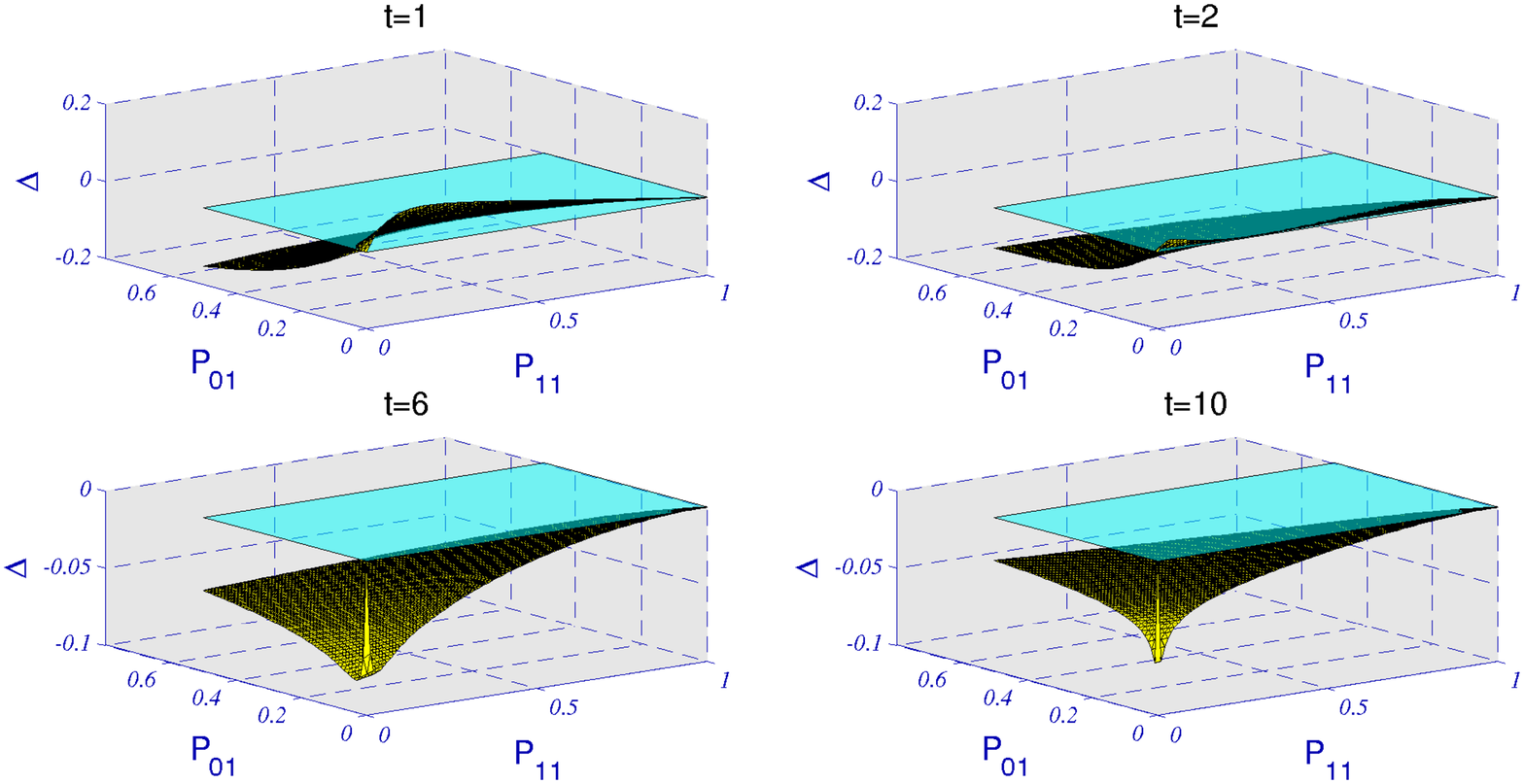}{13.5cm}{Value of $\Delta$ at different generations for a two-locus 
two-allele system with a fitness landscape with genetic redundancy ($b_1=b_2=1$, $c=-1$, $a=1$) 
for different values of the initial population given by $P_{11}$ and $P_{10}(=P_{01})$. The 
$\Delta=0$ plane has been marked to distinguish between conditions in which recombination 
is favorable ($\Delta<0$) or not.}{ORDelta.eps}

\subsubsection{Deceptive Landscape, $a=1$, $b=-0.5$, $c=2$}

Finally, a deceptive landscape (Figure \ref{DeceivingDelta.eps}) offers a complete contrast to that of a redundant one, with recombination being
disadvantageous  in terms of $\Delta$ for any initial population.

\ImagehereFull{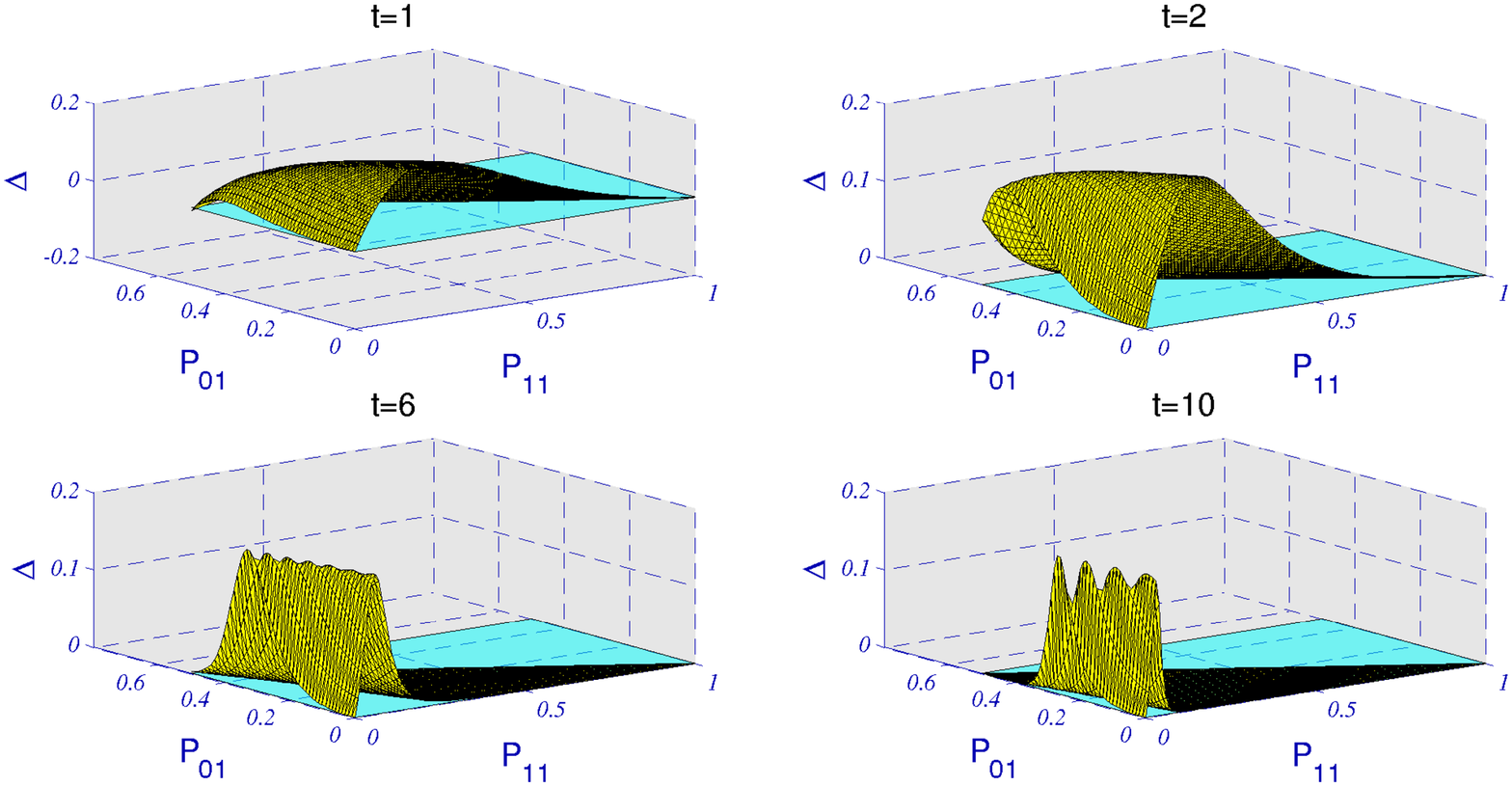}{13.5cm}{Value of $\Delta$ at different generations for a two-locus 
two-allele system with a deceptive fitness landscape ($b_1=b_2=-0.5$, $c=2$, $a=1$) 
for different values of the initial population given by $P_{11}$ and $P_{10}(=P_{01})$. The 
$\Delta=0$ plane has been marked to distinguish between conditions in which recombination 
is favorable ($\Delta<0$) or not.}{DeceivingDelta.eps}

\section{Conclusion}
\label{conclusion}

As discussed in the introduction, genetic recombination remains a puzzle as far as having 
a full, intuitive understanding of why it is so prevalent, with no generally accepted explanation 
of its benefits. Many theoretical analyses have been performed. The vast majority of these 
have been in the context of variations on a theme of standard population genetics models - 
haploid, diploid, with modifer genes, without modifier genes, with finite population, with 
infinite population, with mutation, without mutation, with few loci, with many loci, with 
different fitness landscapes, with different population states etc. Of course, to understand 
the benefits of recombination in the context of a mathematical model, the model itself 
must contain a description of the mechanisms that explain why it is useful in the first place. The 
question is then: do the benefits lie outside the context of the models that have been studied,
or are they hidden within the results of these models? If the former is true, then one must formulate 
a new model, with new features, which will then make manifest its utility. On the other hand, if the latter is the 
case, then it is important to have a model that can be studied exhaustively, in that there is no region 
of the parameter space of the model that remains unexplored. Additionally, the model should 
be such that the effective degrees of freedom of the underlying system are manifest. 

Previous work \cite{stephens2007just,stephens_cervantes}, both analytical and numerical, has hinted at the fact 
that recombination seems to be especially useful in the context of quasi-additive landscapes, while other
work has shown a role for weak negative multiplicative epistasis. However,
these analyses did not cover the full parameter space of the considered models, and so there is 
always doubt that the landscapes or initial populations considered were not 
representative and therefore any identified benefits of recombination were not ``universal" but, rather,
tied to the specific scenario considered. To counter these arguments, in this paper, we have taken 
the route of fixing a simple model - a two locus, two allele system of haploid sequences with 
non-overlapping generations evolving in the presence of selection and homologous recombination - 
but have analyzed the full parameter space of the model. This corresponds to three population 
variables and three landscape parameters. Having fixed the model, we can begin to look for 
the regions of parameter space, if any, in which recombination is beneficial. Of course, we first have
to define what we mean by ``beneficial''. In this paper we fixed two metrics: one was the SWLD 
coefficient for the optimal genotype that measures the excess production of such types over and
above that which is produced by selection only; and the other is the increase in average population
fitness over and above that which would be produced by selection only. With these two metrics we 
measure the benefits of recombination in terms of its capacity to lead to higher proportions of 
fitter genotypes and fitter populations relative to selection only.

So, what does our analysis of the parameter space of this model tell us? The analyses we have 
carried out are consistent with the previous results of \cite{stephens_cervantes} where it was shown that 
there are two important, but distinct, regimes in which recombination is beneficial in terms of both the metrics
that we have used to characterize its benefits. The first of these is the {\it search} regime, which 
is associated with conditions where the fittest genotype is either not present or only at low frequency.
In this regime the benefit from recombination is relatively independent of the fitness landscape. 
However, exactly how beneficial it is does depend on both the landscape and the actual population. 
The second regime we have termed the {\it modular} regime and is associated with weakly additively 
epistatic landscapes, i.e., quasi-additive landscapes. However, the fact that we have analyzed the 
set of possible landscapes and populations, allows us to go beyond this restricted analysis and observe and characterize several important 
universal properties of recombination. 

Firstly, in terms of $\Delta$ there is a clear association
between the sign of the epistasis and the sign of $\Delta$. Production of the optimal genotype ($11$) is 
more favorable in the presence of negative additive epistasis ($c<0$) than for positive additive 
epistasis ($c>0$) for beneficial mutations. It is also disfavored when single mutants ($01$ and $10$) are 
less fit ($b<0$) than the suboptimal genotype $00$. What is more, by following the dynamics across multiple 
generations, we see that recombinative evolution itself is directed towards favoring landscapes that
are more and more modular, more and more negatively epistatic. This is a universal feature that
is independent of the initial population. 

In terms of the increase in average population fitness relative to selection only dynamics we see 
a profoundly interesting dynamic. For the different initial populations considered when investigating 
evolution as a function of landscape, we see that there is an initial regime (t=1) wherein there is a perceived
benefit from recombination for a wide array of landscapes with, in fact, under some circumstances,
a relative advantage for landscapes with positive versus negative additive epistasis. However, 
as evolution progresses, $t=10$, the benefit from recombination has become restricted to 
quasi-additive or negatively additively epistatic landscapes independently of the initial population.  
This is best understood by viewing Figure \ref{FitnessHomogeneous.eps}, where the initial population 
is homogeneous, which means that it begins on the Geiringer manifold. There, we see that recombination
is disfavored initially ($t=1$) for any positively {\it multiplicatively} epistatic landscape - including deceptive
landscapes - and for any {\it additively} negatively epistatic landscape. However, very quickly the 
universal tendency towards favoring quasi-additive and negatively additively epistatic landscapes sets in. There are many works \cite{azevedo2006sexual,burch2004epistasis,maccarthy2007coevolution}, some recent \cite{khan2011negative,kryazhimskiy2011evolution}, in which the role of negative epistasis between 
mutations in evolution is discussed. It must be noted that in these references negative epistasis means sub-multiplicative epistasis, that is, epistasis is quantified with a parameter 
 whose magnitude measures deviations of the logarithm of fitness from linearity as a function of the number of mutations. In our results we included both supra and sub-additive (concerning the sign of $c$) and supra and sub-multiplicative (concerning the sign of $ac-b^2$) epistatic regimes and, importantly it is the
existence of negative additive epistasis that seems to be important for recombination.

As a function of the initial population, we see a complementary but completely consistent point of view 
relative to that of landscape. At $t=1$ we can see the effect of any initial linkage disequilibrium with the 
sign of $\Delta$ being strongly affected by the sign of $D$: more/less BBs relative to $11$ or $00$ being associated
with $D<0$/$D>0$. The effect of deception is to disfavor recombination for basically any population, while
for a genetically redundant landscape it is to favor it for any initial population. 

We believe that the results of this paper unite various important threads of modern evolutionary thought - 
the ubiquity of genetic recombination, the ubiquity of modularity and, relatedly, the ubiquity of
genetic redundancy, and thereby offer a quite universal explanation of why recombination is so widespread. 
This paper is not the appropriate forum in which to discuss the reasons why modularity and redundancy 
themselves are so important. There are many papers on the subject. However, it is amazing that the benefits of 
recombination seem to be so intimately tied to these phenomenon, at least in the framework of the 
fitness landscape paradigm as discussed here. In the space of all possible landscapes we have shown that 
the benefits of recombination are manifest only for quasi-additive or negatively additively epistatic landscapes, a quite
restricted subset of landscape space. However, it is precisely such landscapes that seem to be so common in 
biology. In other words our conclusion is that recombination is so widespread
because it leads to important evolutionary benefits only for systems that are modular and/or redundant and
and it is precisely such landscapes that seem to be the norm. This leads, indeed, to another evolutionary ``chicken 
and egg'' puzzle. Did recombination evolve to take advantage of the existence of modularity and 
redundancy or vice versa? We would posit that there has been a strong co-evolutionary link between 
the them since the beginnings of life with recombination distributions and fitness landscapes co-evolving 
to maximize the benefits of one with the other.

So, what are weak points of our model and analysis? Well, first of all one could criticize the 
simplicity of the model, although the model shares many features with previous analyses. The
fact that only two loci are considered is the price we pay for being able to consider the full parameter
space. However, its worth mentioning again that these ``loci'' could represent different levels
of description from, in principle, nucleotides up to entire sets of genes. Our other restriction is that we
can describe each locus in terms of two possible states. We are quite sure that no qualitative 
effect that we have observed here depends on the existence of only two alleles. The question 
is: are the effects we see and the conclusions we make from the two locus model generalizable 
to multi-loci models? Unfortunately, we cannot analyze exhaustively the full parameter space 
of such a model. For $\ell$ loci there are, in principle, $2^\ell-1$ population parameters and
$2^\ell$ landscape parameters to contend with. 

However, there are some related analyses with multiple  loci \cite{rosenblueth2009analysis}, investigating numerically the dynamics for certain specific landscapes and initial 
populations. The results seen there are completely consistent with what we observe in full generality 
in this paper, i.e., that the benefits of recombination when not in the {\it search} regime are manifest
in modular landscapes while, on the contrary, it is detrimental in the presence of high positive epistasis. 
In this paper we have also neglected the effects of mutation, whereas much previous work has 
been associated with studying how recombination interacts with mutation by positing Muller's 
ratchet type regimes where the dynamics of beneficial or detrimental single mutations are considered
in the presence of recombination. It is an important question to understand the relative benefits
of mutation versus recombination in the context of the metrics that we have considered here. We
will, indeed, return to that in a separate paper. However, it is first important to understand what
benefits there are that are intrinsic to recombination without a comparison with mutation. 

Finally, we have also restricted attention here to fixed-length sequences. We believe that the relation
between recombination and modularity extends beyond this restriction, applying also to variable-length
sequences and recombination-like genetic operators other than homologous recombination. For instance,
unequal crossing over or gene duplication.

\section{Acknowledgements}
\label{acknow}

This work was partially supported by DGAPA grant IN120509 and by a special
Conacyt  grant to the Centro de Ciencias de la Complejidad. DAR is
grateful to the IIMAS, UNAM for use of their facilities. We are grateful to Le\'on Mart\'\i nez and
Michael Gaunt for discussions.

\bibliographystyle{elsarticle-num}
\bibliography{TBTLBtex.bib}

\end{document}